# Resolving cell states in spatial transcriptomics with SCATTER

Lei Huang[1], Yun Gong[2], Weiqiang Lin[2], Lindong Jiang[2], Hui Shen[2], Chaoyang Zhang[1,*], and Hong-Wen Deng[2,*]

1 School of Computing Sciences and Computer Engineering, The University of Southern Mississippi, Hattiesburg, MS, USA. 2 Deming Department of Medicine and Tulane Center for Biomedical Informatics and Genomics, Tulane University School of Medicine, New Orleans, LA, USA.

*Corresponding authors: Hong-Wen Deng (hdeng2@tulane.edu) and Chaoyang Zhang (chaoyang.zhang@usm.edu).

## Abstract

In spot-based spatial transcriptomics, expression measured at a spot can reflect contributions from multiple cell types. Differences in expression between spots may therefore arise from changes in cell-type proportions, changes in expression within a cell type, or both. These sources of variation are difficult to distinguish from mixed counts alone. We developed SCATTER to estimate cell-type proportions and the relative expression profile of each cell type at each spot from spatial transcriptomic counts, spot coordinates, and annotated single-cell reference data. SCATTER first integrates expression from a spot and its neighbors to jointly estimate cell-type proportions and local expression profiles using softmax reference coupling and a Cell2location proportion prior. It then holds those proportions fixed while estimating local cell-type expression profiles with second-order Taylor reference coupling. Gene calibration and residual backfitting refine the local profiles against the observed counts. Across six pseudo-spot benchmarks, SCATTER yielded higher within-cell-type spatial correlations than DestVI, PANDA and STged for 79.6–98.9% of paired gene–cell type comparisons. These comparisons required at least 20 eligible spots and nonzero target expression in more than 15% of those spots. In the breast cancer Xenium benchmark, SCATTER recovered tumor neighborhood–associated gene expression and regulon scores. In the multiple sclerosis MERFISH benchmark, SCATTER recovered lesion-associated glial expression programs. In an exploratory analysis of a breast cancer Visium section, inferred stromal and macrophage expression showed differing associations with the neighborhood malignant cell fraction. The colorectal cancer Visium HD analysis characterized cell-type-specific expression programs at the invasive front and ranked nine candidate ligand–receptor pairs using inferred expression in the corresponding cell types. By improving the recovery of spatial expression variation within cell types, SCATTER supports analyses of gene expression and regulon score patterns across tissue microenvironments.

## Introduction

Gene expression within the same cell type can vary across tissue microenvironments[1,2]; for example, macrophages near a tumor may show different expression patterns from macrophages farther away. Spatial transcriptomic platforms differ in the size of their capture areas[3,4]. In conventional Visium, a spot typically captures transcripts from multiple cells. Therefore, if a gene shows higher expression in spots near a tumor, this may reflect either a higher proportion of a cell type that strongly expresses the gene or higher expression of the gene within a particular cell type. Distinguishing these two sources of variation requires estimating both the cell-type composition of each spot and gene expression within each cell type at that spot.

Deconvolution methods such as Cell2location[5], STINR[6], RCTD[7], and CARD[8] use single-cell reference data to estimate the cell-type composition of each spot. Building on this, to infer cell-type-specific expression profiles at each spot, one approach uses pre-estimated proportions as fixed inputs and then fits expression profiles, whereas another estimates proportions and expression profiles jointly. STged[9] follows the first approach. It uses regularized nonnegative least squares to fit expression profiles, penalizing deviations from the cell-type mean reference profile as well as differences between profiles of the same cell type at neighboring spots. Because proportions remain fixed, errors in the pre-estimated proportions can carry over into the fitted expression profiles. In contrast, joint estimation allows proportions to adjust during expression fitting. DestVI[10] and PANDA[11] adopt this approach. DestVI first trains a conditional variational autoencoder on single-cell reference data to learn how to generate expression profiles from cell types and cell states. When fitting spatial counts, it uses a negative binomial count model to jointly estimate cell-type abundances and cell states at each spot and uses the trained decoder to convert those states into expression profiles. PANDA uses probabilistic archetypal analysis to derive expression archetypes from the reference data and then uses a Poisson count model to jointly estimate archetype weights and cell-type abundances at each spot. However, in their standard formulations, both methods infer local expression from the observed counts at each spot without explicitly incorporating counts from neighboring spots[10,11], leaving out contextual signals from adjacent spots within the same local microenvironment.

To address this, we developed SCATTER. The first stage uses a spatial encoder to incorporate count profiles and local tissue context, while softmax matching within annotated reference-cell populations constructs type-specific expression profiles. Cell-type proportions and expression profiles are jointly fitted to observed mixed counts, with composition regularized toward Cell2location estimates. To refine local expression profiles at the estimated composition, SCATTER initializes a second model with the same architecture and fits it to the observed counts with the first-stage proportions held fixed. Mixed counts provide indirect evidence about which reference cells represent a given cell type at a spot, so the second stage uses second-order Taylor weighting to temper the contrast between reference-cell weights for a given set of matching scores. After fitting, gene calibration and residual backfitting refine expression profiles against observed counts.

To assess whether this approach recovers spatial expression variation within a cell type, we constructed six pseudo-spot benchmarks from single-cell spatial expression data generated by MERFISH, Xenium and Stereo-seq. We summed transcript counts within spatial bins and evaluated the recovered profiles against cell-type-specific targets derived from the single-cell data. For each gene–cell type pair, recovery was assessed by correlating predicted and target expression across pseudo-spots. SCATTER achieved the highest mean within-cell-type spatial correlation in each of the six benchmarks. For gene–cell type pairs with at least 20 eligible spots and nonzero target expression in more than 15% of them, SCATTER achieved higher correlations in 79.6–98.9% of paired comparisons with DestVI, PANDA and STged. The breast cancer benchmark also tested recovery of tumor neighborhood–associated expression and regulon-based activity scores. The cross-modality multiple sclerosis benchmark examined recovery of lesion-associated glial expression programs. In directly measured breast cancer Visium data, we examined how stromal and macrophage expression varied with the malignant-cell proportion in neighboring spots. The colorectal cancer Visium HD analysis mapped expression programs at the tumor–stroma interface and ranked candidate ligand–receptor pairs.

# Results

## Overview of SCATTER

SCATTER takes mixed transcript counts, spot coordinates and an annotated single-cell reference as inputs (Fig. 1a). During training, the spot encoder combines log-normalized counts with Fourier features of the coordinates, then uses multi-head attention over the spatial neighborhood to incorporate expression from adjacent spots (Fig. 1b). A multilayer perceptron (MLP) projects the resulting spot representation for use in a spot- and cell-type-specific query. A separate reference encoder maps each reference cell through an MLP and L2-normalizes the resulting embedding. Comparing each query with the embeddings of reference cells of the corresponding type gives weights for their expression profiles. SCATTER sums the weighted reference-cell profiles within each type to obtain a local expression profile.

To reconstruct mixed spot counts, SCATTER combines the local profiles weighted by their cell-type proportions. In both stages, learned per-gene factors scale the reconstructed counts before the count reconstruction loss compares them with observed counts. Stage 1 uses softmax reference-cell weights and jointly fits the proportions and local profiles. A Kullback–Leibler (KL) penalty regularizes the proportions toward Cell2location estimates. Stage 2 starts from a fresh initialization of the same architecture, holds the Stage 1 proportions fixed and uses second-order Taylor weights to fit local expression profiles.

At inference, SCATTER applies the learned Stage 2 gene factors to each local profile and renormalizes it across genes. Each gene factor is shared across spots and cell types. Residual backfitting keeps Stage 1 proportions fixed. For each type in turn, it subtracts the estimated contributions of the other types from the observed spot expression rates and blends a residual-based estimate into the current profile. The two outputs are the spot-level proportions and the refined relative expression profile for each cell type at each spot (Fig. 1c). These outputs support mapping abundance and expression, assessing within-cell-type spatial heterogeneity and ranking cell-type markers (Fig. 1d).

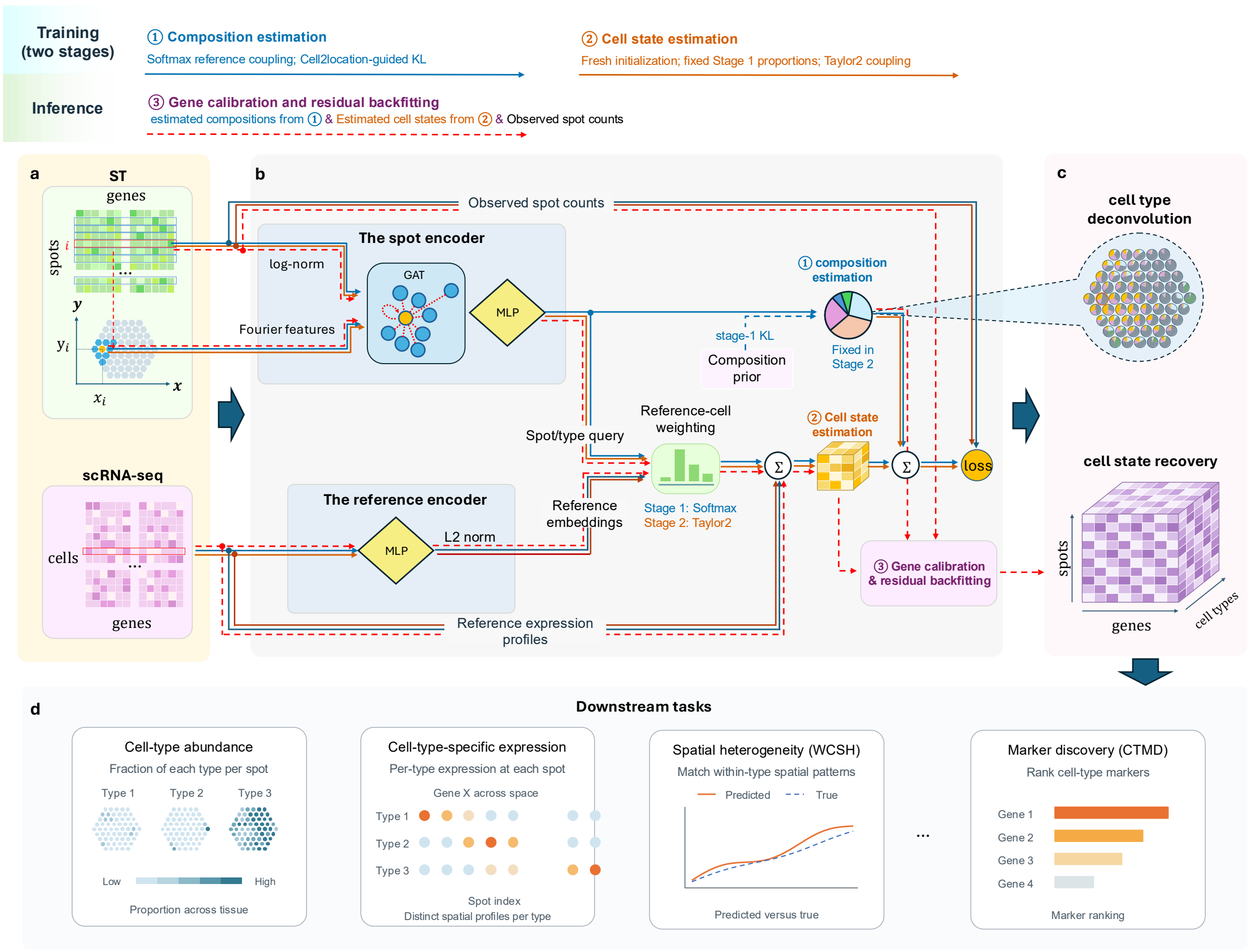

**Fig. 1. Overview of SCATTER.** a, Inputs comprise spot-by-gene counts, spatial coordinates and an annotated single-cell reference. b, SCATTER uses two sequential fitting stages (steps 1 and 2) with the same architecture but independently optimized parameters. A spatial graph-transformer encoder integrates normalized counts and Fourier coordinate features, while a reference encoder generates normalized reference embeddings. Within each annotated cell type, softmax coupling in Stage 1 or second-order Taylor (Taylor2) coupling in Stage 2 weights reference cells to construct cell-type-specific expression profiles (first Σ). A second Σ combines these profiles across cell types using spot-specific proportions to reconstruct mixed counts for training. Stage 1 estimates proportions under Cell2location-guided Kullback–Leibler (KL) regularization. Stage 2 starts from a fresh initialization and fits expression profiles with Stage 1 proportions fixed. During inference (step 3), Stage 1 predicts proportions and Stage 2 estimates expression profiles, which undergo gene calibration and residual backfitting against observed counts with proportions fixed. Blue and orange solid paths indicate Stage 1 and Stage 2, respectively; red dashed paths indicate inference. The gold tensor represents initial expression profiles. c, Outputs comprise cell-type proportions and the final spot-by-cell-type-by-gene expression tensor, shown in lavender. d, Applications include cell-type abundance and expression mapping, within-cell-type spatial heterogeneity (WCSH) assessment and cell-type marker discovery (CTMD).

## Benchmarking cell-type proportions and spatial expression across platforms

We selected spatial datasets to test whether SCATTER could recover cell-type proportions and expression patterns from mixtures constructed from imaging or sequencing data, and to assess its estimates on directly captured tissue data. In the MERFISH and Xenium pseudo-spots, measurements from the constituent cells provided targets for evaluating recovery. The Stereo-seq benchmark extended this test to sequencing-derived mixtures and used separate transcript pools from the constituent cells to construct the mixed counts and expression targets (Table 1 and Supplementary Fig. 5). We then assessed predictions from directly captured lung Visium HD counts against Xenium measurements from the same or adjacent sections (Table 2). Marker-gene concordance provided an indirect assessment of cell-type proportions in melanoma and cutaneous squamous cell carcinoma sections. The sci-Space pseudo-spot benchmark tested recovery with single-cell references drawn from the constituent cells or from held-out sections (Supplementary Table 10).

## Primary pseudo-spot benchmarks from MERFISH, Xenium and Stereo-seq

We constructed six pseudo-spot benchmarks from spatial transcriptomic data with single-cell resolution. The datasets comprised a coronal section from a whole-brain MERFISH dataset[12], MERFISH data from the hypothalamic preoptic area[13], human breast cancer Xenium data[14], and MERFISH data from a human multiple sclerosis (MS) brain tissue cohort[15]. The MS cohort contributed two settings, one with a cross-modality single-nucleus reference and one with a same-modality reference constructed from MERFISH cells excluded from pseudo-spot construction. The sixth benchmark used Stereo-seq cell-bin data from an adult mouse hemibrain[16]. In all six benchmarks, we summed cell-level counts within spatial bins. The five MERFISH and Xenium settings used the same single-cell measurements to construct mixed counts and cell-type-specific expression targets. In Stereo-seq, we randomly assigned each cell's transcripts to two pools. One pool supplied the mixed counts and the other supplied the expression targets. The breast cancer benchmark used an independent single-cell scFFPE-seq reference from formalin-fixed, paraffin-embedded (FFPE) tissue. Evaluation was restricted to the 313 genes shared between the reference and the imaging assay.

Within-cell-type spatial correlation measures how closely predicted and target expression for a gene–cell type pair vary across space. For each pair, we calculated Pearson correlation between predicted and target expression across eligible pseudo-spots, defined as spots in which the true proportion of that cell type exceeded 0.05. We compared SCATTER with DestVI, PANDA, and STged, all of which provide spot-level expression profiles for each annotated cell type. All methods were evaluated on the same gene set within each dataset. Mean correlations were calculated over pairs with at least three eligible spots and a defined correlation, without filtering on expression frequency. For each paired baseline comparison, we required at least 20 eligible spots, nonzero target expression in more than 15% of those spots, and a defined correlation for both methods (Supplementary Note 2).

SCATTER achieved mean within-cell-type spatial correlations of 0.233–0.356 across the six benchmarks, exceeding the best-performing baseline in each benchmark. The corresponding mean correlations for the best-performing baselines ranged from 0.094 to 0.149 (Table 1 and Supplementary Table 2). In paired comparisons with DestVI, PANDA and STged, SCATTER achieved higher correlations for 79.6–98.9% of the evaluated pairs (Table 1 and Supplementary Table 3). In the Stereo-seq benchmark, the corresponding win rates against DestVI, PANDA and STged were 96.3%, 96.1% and 81.0%, respectively (Supplementary Table 3). All 18 method comparisons across the six benchmarks were significant after one-sided paired Wilcoxon signed-rank tests and Benjamini–Hochberg correction ($q < 10^{-5}$). The Stereo-seq scatter plots compare all three baselines using at least 30 eligible spots and no expression-frequency filter. In that analysis, SCATTER had a higher correlation than STged for 3,779 of 6,431 pairs (58.8%, Supplementary Fig. 5). The whole-brain MERFISH maps illustrate the recovered cell-type-specific expression patterns for 12 cell types and the reconstructed mixed-spot expression (Supplementary Fig. 2).

**Table 1. Cell-type proportion and expression recovery in six primary pseudo-spot benchmarks.**

| Tissue | Spatial platform | Reference data | Proportion PCC | Mean within-type spatial PCC | Paired expression win rate |
|---|---|---|---|---|---|
| Mouse whole brain | MERFISH | 10x scRNA-seq | **0.890** / 0.872 (Cell2location) | **0.356** / 0.143 (PANDA) | 80.6–84.4% |
| Mouse hypothalamus | MERFISH | Moffitt scRNA-seq | 0.866 / **0.872** (Cell2location) | **0.290** / 0.116 (DestVI) | 79.6–85.2% |
| Human breast cancer | Xenium | scFFPE-seq | 0.730 / **0.736** (DestVI) | **0.287** / 0.094 (STged) | 88.5–92.8% |
| Human MS brain | MERFISH | snRNA-seq | **0.858** / 0.853 (Cell2location) | **0.276** / 0.147 (STged) | 96.9–98.9% |
| Human MS brain | MERFISH | MERFISH | **0.887** / 0.883 (Cell2location) | **0.274** / 0.149 (STged) | 91.0–97.1% |
| Mouse hemibrain | Stereo-seq | 10x scRNA-seq | 0.643 / **0.646** (Cell2location) | **0.233** / 0.139 (STged) | 81.0–96.3% |

Note. For each benchmark, proportion PCC is shown for SCATTER and the highest-scoring proportion baseline. Mean within-type spatial PCC is shown for SCATTER and the highest-scoring expression baseline among DestVI, PANDA and STged. Baseline names are in parentheses. Proportion PCC is averaged across cell types. For each method, mean within-type spatial PCC is averaged over gene–cell type pairs with a defined Pearson correlation and at least three spots in which the true cell-type proportion exceeded 0.05. The win-rate range spans the fraction of matched pairs on which SCATTER had a higher correlation than each of the three expression baselines. For each method comparison, pairs required at least 20 such spots, nonzero target expression in more than 15% of those spots and a defined correlation for both methods (Supplementary Note 2). MS denotes multiple sclerosis.

Figure 2 illustrates recovery in breast cancer Xenium pseudo-spots, where the imaging-derived expression targets allow the spatial maps to be evaluated before examining their associations with the tumor neighborhood. The breast cancer Xenium benchmark used an scFFPE-seq reference from the same tissue block with no overlapping cells to evaluate recovery across assays. Within each spot, a single Pearson correlation compared predicted and imaging-derived expression values concatenated across genes and cell types represented by at least one constituent cell. The mean within-spot correlation was the average of these correlations across spots. SCATTER achieved a mean within-spot Pearson correlation of 0.779, compared with 0.724 for the best-performing baseline (Supplementary Table 2). In the analysis across spots, SCATTER achieved a higher Pearson correlation than DestVI for 88.5% of 3,982 evaluated gene–cell type pairs and a higher Spearman correlation for 88.5% of 4,025 evaluated pairs (Supplementary Table 3). The expression maps show that SCATTER retained regional differences in ERBB2 expression in invasive tumor cells and CCND1 expression in the DCIS 2 population. Several baseline maps showed smaller differences between regions (Fig. 2). To illustrate mixed-spot reconstruction separately from type-specific recovery, Supplementary Fig. 1 shows reconstructed expression maps for ADIPOQ, LEP, MS4A1, and CTSG. The four examples in Fig. 2 were selected for biological relevance, coverage of imaging-derived target expression and higher SCATTER recovery correlations.

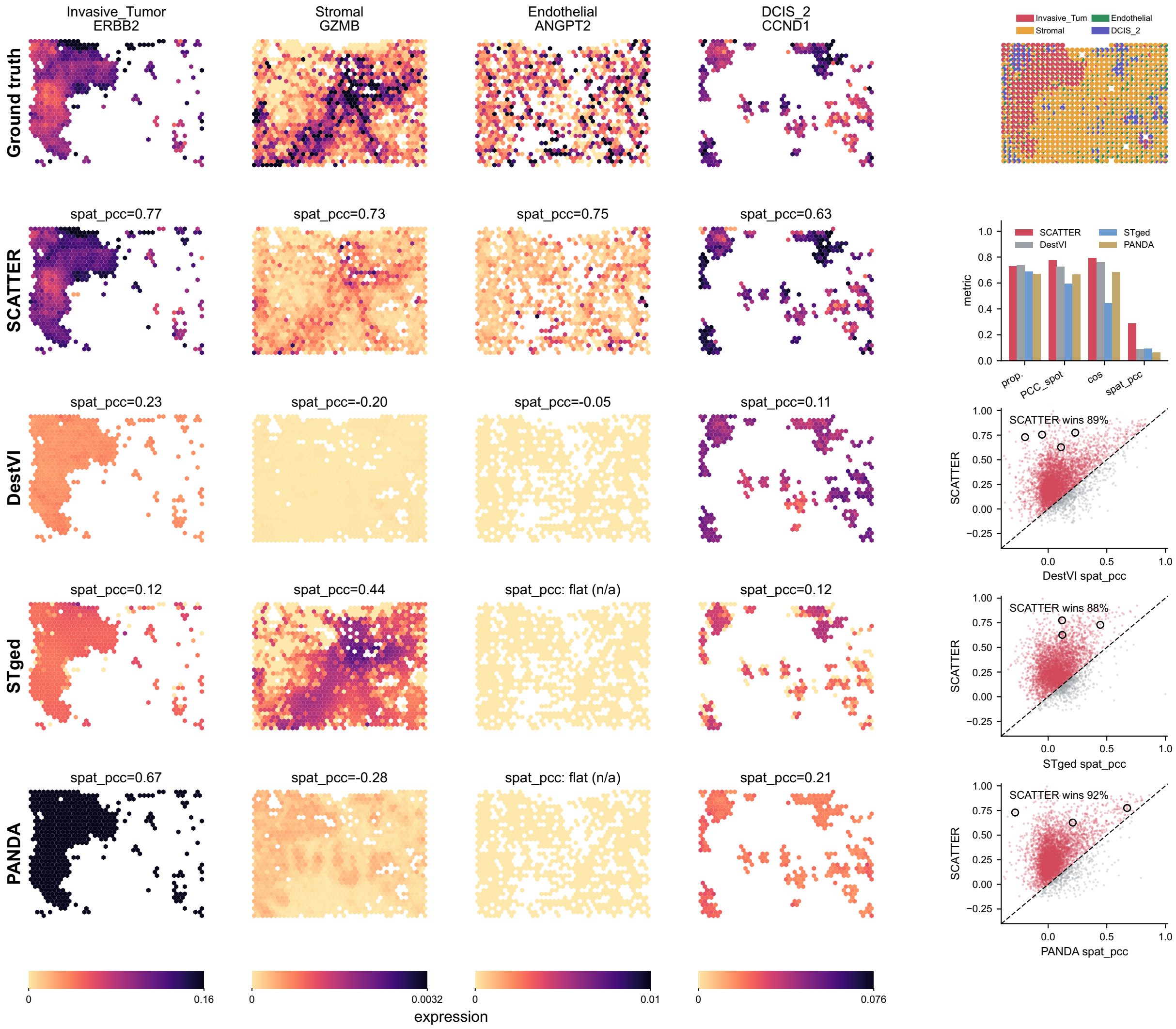

**Fig. 2. Recovery of cell-type-specific expression in pseudo-spot benchmarks.** Spatial maps show ERBB2 in invasive tumor cells, GZMB in stromal cells, ANGPT2 in endothelial cells and CCND1 in the ductal carcinoma in situ population DCIS 2 in breast cancer Xenium pseudo-spots. Rows show imaging-derived targets and predictions from SCATTER, DestVI, STged and PANDA. Maps in each column share a color scale. Labels give within-cell-type spatial Pearson correlations, with n/a indicating an undefined correlation. Pie charts show the proportions of the four displayed cell types, renormalized to sum to 100%. Bars compare proportion PCC, within-spot expression PCC, profile cosine similarity and mean within-cell-type spatial PCC. Each scatter plot compares SCATTER with one baseline across gene–cell type pairs with defined correlations and at least 30 spots where the true proportion exceeded 0.05. Red points favor SCATTER, open circles mark the mapped examples and dashed diagonals indicate equality. Table 1 reports win rates under the paired-test criteria in Supplementary Note 2.

For each gene–cell type pair, we calculated Moran's I separately for predicted and target expression. The correlation between the resulting Moran's I values across pairs quantified agreement in spatial autocorrelation. SCATTER showed the highest agreement in all six benchmarks. After averaging across the three overlapping cell-type proportion subsets (all present, non-dominant and dominant), the correlations were 0.67–0.84 for SCATTER and 0.41–0.60 for the best-performing baseline in each benchmark (Supplementary Table 4 and Supplementary Fig. 3).

SCATTER uses the fitted cell-type proportions as fixed inputs when estimating expression profiles. We evaluated these proportions against the true proportions in the six benchmarks and compared SCATTER with nonnegative least squares (NNLS), STINR[6], Cell2location[5], DestVI[10], PANDA[11], and STged[9].We assessed agreement with the true proportions using cell-type-wise Pearson correlations and root mean square error (RMSE). SCATTER achieved the highest cell-type-wise Pearson correlations in the whole-brain MERFISH benchmark and both MS benchmarks, with values of 0.890, 0.858, and 0.887, respectively (Table 1 and Supplementary Table 1). In the hypothalamic MERFISH dataset, SCATTER achieved a correlation of 0.866, close to 0.872 for Cell2location. In the breast cancer Xenium dataset, the corresponding correlations were 0.730 for SCATTER and 0.736 for DestVI. SCATTER also had the lowest proportion RMSE in the Xenium dataset, with a small difference from DestVI (0.082 versus 0.083, Supplementary Table 1). In the Stereo-seq benchmark, the correlation was 0.643 for SCATTER and 0.646 for Cell2location, and SCATTER had the lowest RMSE (0.100).

### Benchmarking in directly measured tumor sections

In directly measured tumor sections, we assessed both cell-type proportions and expression in lung and used marker-gene concordance to assess proportions in melanoma and cutaneous squamous cell carcinoma. For the lung benchmark, we compared predictions from three Visium HD lung adenocarcinoma sections with Xenium measurements from the same tissue block. One section was assayed with Xenium v1 before Visium HD. Two sections were assayed with Visium HD only and compared with adjacent sections measured using Xenium v1 or Xenium Prime 5K. On each section, SCATTER had the lowest cell-type proportion RMSE among the seven methods evaluated. Its mean within-cell-type spatial PCC against Xenium profiles was higher than those of DestVI, PANDA and STged on each section (Table 2, Supplementary Note 4 and Supplementary Table 9).

**Table 2. Cell-type proportion and spatial expression estimates on lung Visium HD sections evaluated against Xenium.**

| Visium HD workflow | Section pairing | Xenium panel | Genes scored | Proportion RMSE | Mean within-type spatial PCC |
|---|---|---|---|---|---|
| After Xenium | Same section | v1 | 175 | **0.085** / 0.097 (PANDA) | **0.170** / 0.095 (STged) |
| Visium HD only | Adjacent section | v1 | 175 | **0.092** / 0.101 (Cell2location) | **0.171** / 0.094 (STged) |
| Visium HD only | Adjacent section | Prime 5K | 4,441 | **0.107** / 0.116 (Cell2location) | **0.055** / 0.029 (STged) |

Note. The section labeled "After Xenium" was assayed with Xenium v1 before Visium HD. For the two "Visium HD only" sections, the indicated Xenium panels were applied to adjacent sections from the same tissue block. Proportion RMSE is shown for SCATTER and the lowest-RMSE baseline among the evaluated proportion methods in each section. Mean within-type spatial PCC is shown for SCATTER and the highest-scoring expression baseline (STged in all three sections). It was calculated on gene–cell type pairs with a defined correlation for all expression-recovery methods, excluding the genes used for Xenium cell-type annotation. Xenium cell-type fractions and whole-cell expression profiles were computed in registered 128-μm squares. For each cell type, expression was scored in squares whose Xenium cell-type fraction exceeded 0.05. Supplementary Table 9 gives results for all expression baselines and paired win rates.

Marker-gene concordance provided an additional assessment of cell-type proportions in spatial data measured directly from melanoma[17] and cutaneous squamous cell carcinoma[18] sections (Supplementary Table 5). In melanoma, SCATTER showed the highest concordance with curated marker-gene programs (0.639), close to the score for its Cell2location prior (0.634). In cutaneous squamous cell carcinoma, SCATTER scored 0.461, and DestVI had the highest score on that section (0.500).

### sci-Space embryonic cortex pseudo-spots

We also evaluated SCATTER on the sci-Space mouse embryonic cortex data used in the DestVI study[10]. We evaluated the original mixtures and mixtures with increased contributions from non-neuronal cells. For each mixture setting, we used a reference drawn from the constituent cells and a separate reference drawn from held-out sections. On gene–cell type pairs with defined correlations for all four expression methods, SCATTER had a higher mean within-cell-type spatial PCC than DestVI, PANDA and STged in all four settings. PANDA had the highest PCC for cell-type proportions among the seven methods evaluated in each setting (Supplementary Note 4 and Supplementary Table 10).

## Recovery and analysis of tissue-associated expression patterns

We next examined expression patterns associated with specific tissue features. In breast cancer Xenium and multiple sclerosis MERFISH pseudo-spots, targets derived from the constituent cells allowed us to assess recovery of tumor-neighborhood expression associations, regulon scores and lesion-associated glial programs (Figs. 3 and 5). In directly measured breast cancer Visium and colorectal cancer Visium HD sections, we used inferred expression profiles to examine tumor neighborhoods and the invasive tumor–stroma interface (Figs. 4 and 6).

### Recovery of neighborhood-associated expression and regulon scores in breast cancer Xenium pseudo-spots

Using breast cancer Xenium pseudo-spots, we tested recovery of tumor-neighborhood expression associations and regulon-score patterns (Fig. 3). To assess expression associations, we compared how inferred and imaging-derived expression related to the neighborhood malignant cell fraction. For each pseudo-spot, this fraction was the mean of the imaging-derived malignant epithelial cell proportions in the 15 nearest other pseudo-spots, based on the Xenium cell-type annotations. The same fraction was used in both association analyses and withheld during model fitting.

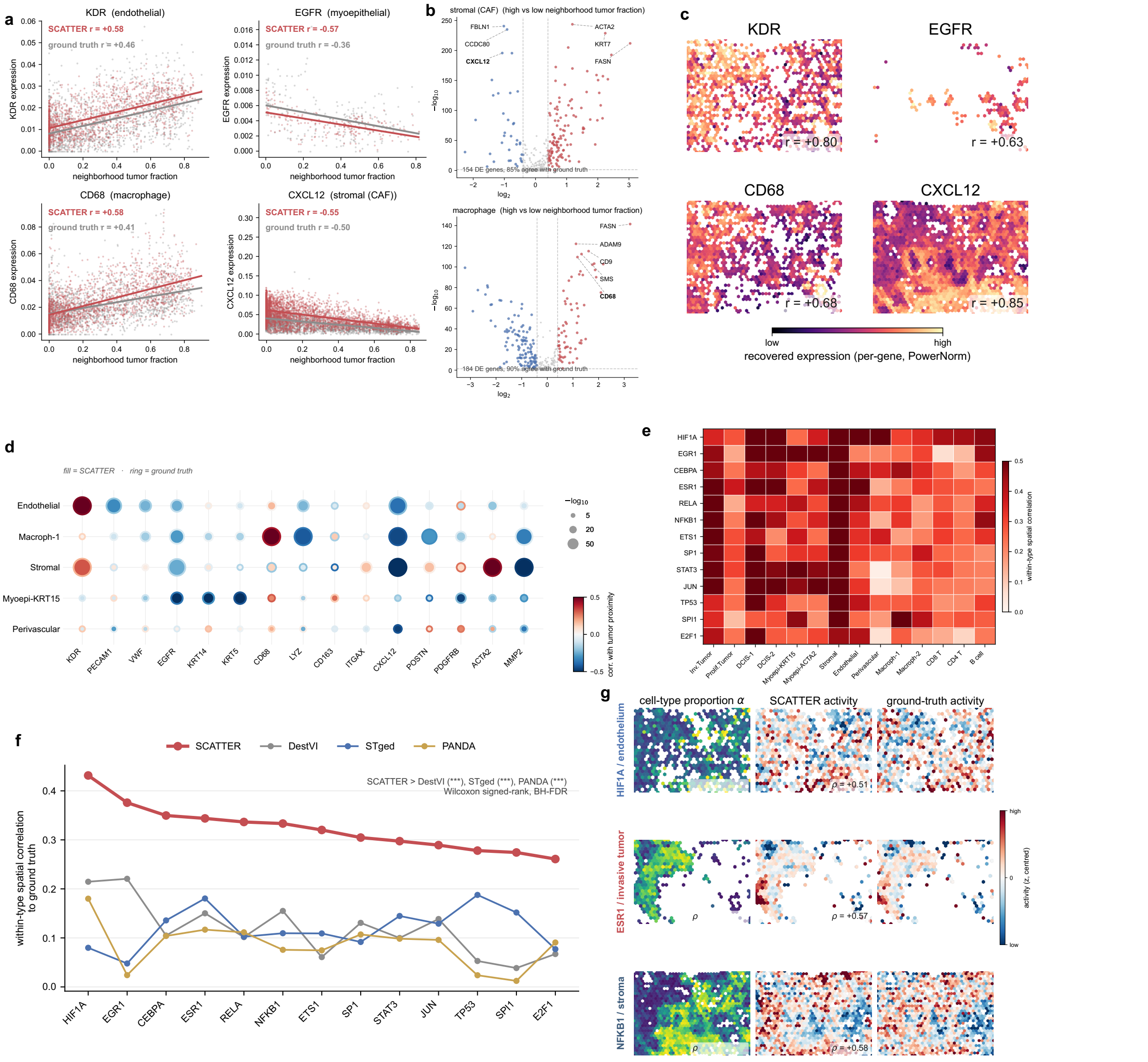

**Fig. 3. Recovery of tumor-neighborhood expression and regulon scores in breast cancer Xenium pseudo-spots.** a, Each point is a pseudo-spot, showing inferred gene expression versus the mean malignant-cell proportion in its 15 nearest pseudo-spots; lines are linear fits and r denotes Pearson correlation. b, Stromal and macrophage expression contrasts between the highest and lowest thirds of neighborhood malignant-cell proportion. Red and blue mark inferred increases and decreases at nominal $P < 0.05$ and absolute $\log_2$ fold change > 0.4; labels indicate whether the direction agrees with imaging-derived targets. c, Inferred and imaging-derived target expression maps with gene-specific scales; r denotes their spatial Pearson correlation. d, Correlations between expression of 15 genes and neighborhood malignant-cell proportion in five cell types. Circle fill and outline represent inferred and target correlations, respectively; area represents the capped $-\log_{10}$ P of the inferred association. e, Within-cell-type spatial Spearman correlations between inferred and target regulon scores for 13 factors across 14 cell types. f, Median recovery correlation across cell types for each factor. Asterisks mark one-sided paired Wilcoxon tests of higher SCATTER recovery across 182 shared factor–cell type pairs, corrected over three baseline comparisons (***$q < 0.001$). g, Imaging-derived cell-type proportions, inferred regulon scores and target scores for HIF1A in endothelial cells, ESR1 in invasive tumor cells and NFKB1 in stromal cells. Scores were centered and scaled separately; ρ denotes the correlation of inferred and target scores; $\rho(A, \alpha)$ relates the inferred score A to the imaging-derived cell-type proportion α. CAF denotes cancer-associated fibroblasts. DCIS-1 and DCIS-2 denote the two ductal carcinoma in situ populations in the source data.

For all four example gene–cell type pairs, inferred and imaging-derived expression were associated with the neighborhood malignant cell fraction in the same direction (Fig. 3a). Endothelial KDR, which is associated with vascular expression programs in breast cancer[19,20], was positively correlated with the neighborhood malignant cell fraction (r = 0.58 for inferred expression and 0.46 for target expression). In myoepithelial cells, which maintain mammary tissue architecture[21,22], EGFR expression was negatively correlated with the neighborhood malignant cell fraction (r = −0.57 for inferred expression and −0.36 for target expression). Expression of the macrophage marker CD68[23] was higher in neighborhoods enriched for malignant cells (r = 0.58 for inferred expression and 0.41 for target expression). Stromal CXCL12, which is associated with fibroblast-related signaling in breast cancer[24], showed a negative association, opposite in direction to CD68 (r = −0.55 for inferred expression and −0.50 for target expression).

The differential expression analysis compared pseudo-spots in the highest and lowest tertiles of neighborhood malignant cell fraction. Among genes identified as differentially expressed using the inferred profiles, the direction of change agreed with imaging-derived target expression for 131 of 154 stromal genes (85%) and 166 of 184 macrophage genes (90%) (Fig. 3b). Stromal ACTA2 expression was higher and CXCL12 expression was lower in the highest tertile than in the lowest tertile. In a breast cancer atlas, inflammatory-like fibroblasts expressed CXCL12, whereas myofibroblast-like cells expressed ACTA2 and were enriched in invasive regions[25]. For the four illustrative spatial expression maps in Fig. 3c, correlations between predicted expression and the targets derived from the constituent cells ranged from r = 0.63 to 0.85. In a separate assessment of 75 gene–cell type pairs, inferred and imaging-derived expression showed associations with neighborhood malignant cell fraction in the same direction for 57 pairs (Fig. 3d).

Regulon scores computed from measured target-gene expression served as benchmarks for scores computed from inferred expression. The analysis used human DoRothEA regulons obtained through OmniPath[26] with at least five measured target genes. Each score was a weighted mean across target genes, using signed regulon weights and expression values standardized across spots for each gene. The comparison of inferred and imaging-derived regulon scores across spots included 182 eligible transcription factor–cell type pairs representing 13 factors. The median Spearman correlation across these pairs was 0.31 for SCATTER. The corresponding medians were 0.12 for DestVI, 0.13 for STged, and 0.08 for PANDA (Fig. 3e,f). Each comparison was significant after one-sided Wilcoxon signed-rank testing and Benjamini–Hochberg correction ($q < 0.001$).

For HIF1A in endothelial cells, ESR1 in invasive tumor cells, and NFKB1 in stromal cells, the within-cell-type spatial Spearman correlations between inferred scores and scores derived from the corresponding constituent cells were 0.51, 0.57, and 0.58, respectively (Fig. 3g). To assess how closely the inferred scores tracked cell-type abundance, we correlated them with the corresponding cell-type proportions. The correlations were 0.03, −0.08 and −0.05, respectively. These examples illustrate the broader comparison of 182 factor–cell type pairs. Supplementary Method 1 describes the assayed target-gene and cell-type proportion criteria.

### Neighborhood-associated expression in directly measured breast cancer Visium data

The breast cancer microenvironment analysis also included directly measured Visium CytAssist data from the same FFPE tissue block as the Xenium benchmark[14]. The analysis included 4,963 spots and 5,012 modeled genes and used an scFFPE-seq reference from the same tissue block. We analyzed SCATTER expression estimates from three runs with different random seeds. For each focal spot, the neighborhood malignant cell fraction was the mean malignant cell proportion estimated by SCATTER (seed 0) across its 15 nearest spots (Fig. 4b). We used this same neighborhood fraction for all three runs. Cell2location estimates from seed 0 defined eligibility for each cell type (proportion >0.05). The analysis included 3,971 eligible spots for stromal cells, 350 for Macrophage 1, and 3,728 for Macrophage 2 (Fig. 4c). Partial Spearman correlations between inferred expression and the neighborhood malignant cell fraction adjusted for the Cell2location seed-0 estimate of the focal cell type's proportion and the total count across modeled genes. Supplementary Method 2 describes the selection of genes for the stromal and macrophage analyses and the fixed masks used across seeds.

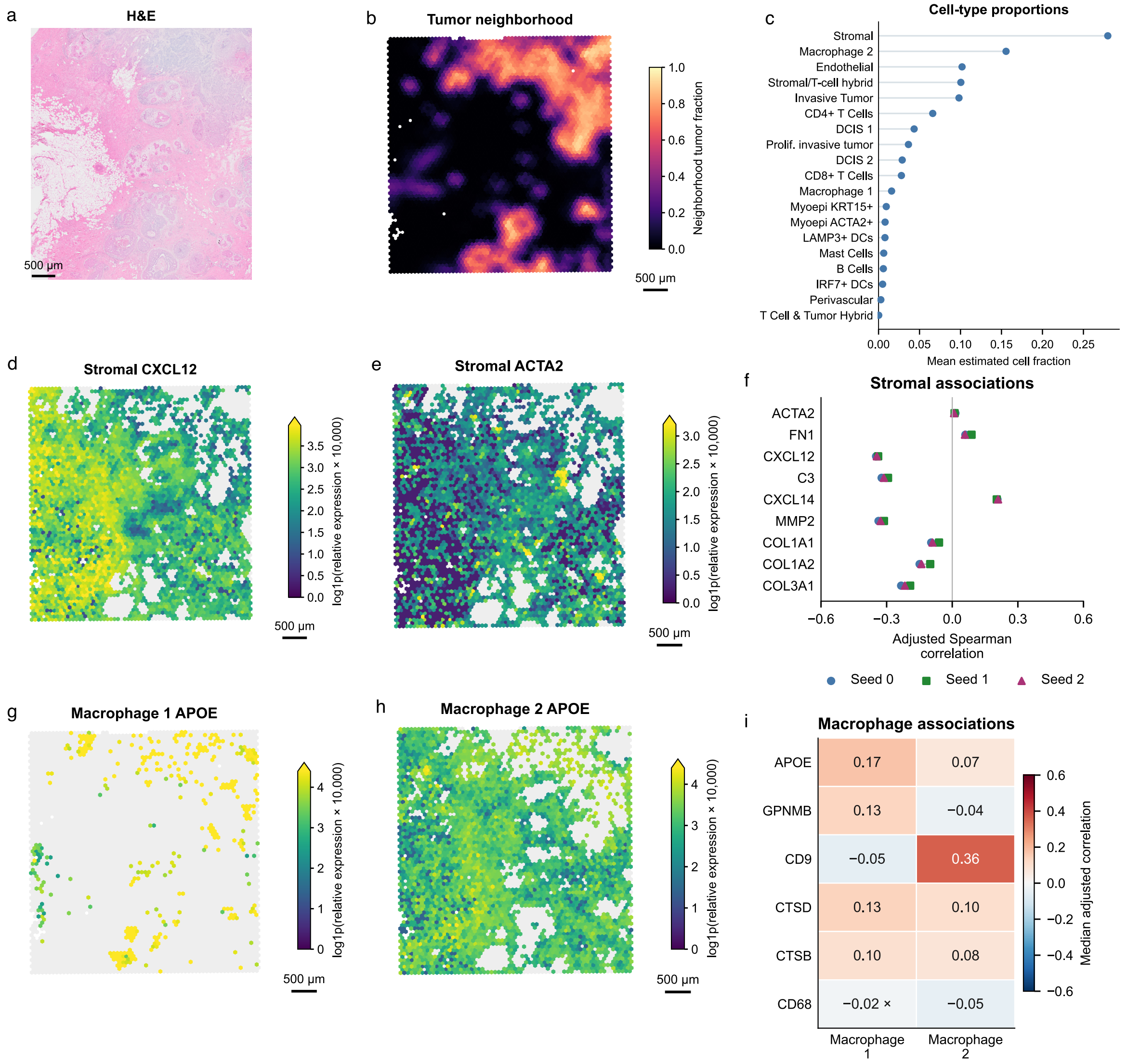

**Fig. 4. Cell type–associated expression across tumor neighborhoods in directly measured breast cancer Visium data.** a, H&E image. b, Mean malignant cell fraction across the 15 nearest spots, inferred by SCATTER using seed 0. c, Mean proportions of 19 cell types across 4,963 spots, estimated by Cell2location using seed 0. d,e, Inferred stromal CXCL12 and ACTA2 expression. f, Partial Spearman correlations between the neighborhood malignant cell fraction and expression of nine stromal genes, adjusted for stromal proportions estimated by Cell2location and library size across the modeled gene panel. Symbols indicate three random seeds. Horizontal lines show the ranges of the estimates across seeds. g,h, APOE expression in Macrophage 1 and Macrophage 2 on a shared color scale. i, Each tile shows the median partial Spearman correlation across three seeds between inferred gene expression and the neighborhood malignant-cell fraction, adjusted for macrophage population proportion and library size. Values in the tiles give the correlations rounded to two decimal places. Red and blue indicate positive and negative associations, respectively. Crosses indicate that the sign of the correlation differed across seeds. Expression maps show the median relative expression across three seeds after a log1p(10,000 × relative expression) transform. Spots with Cell2location proportion ≤0.05 appear in light gray. Color-scale arrows indicate 99th-percentile display limits; the APOE panels share a scale. Scale bars, 500 μm. H&E denotes hematoxylin and eosin, DCIS denotes ductal carcinoma in situ and DCs denotes dendritic cells. Macrophage 1 and Macrophage 2 are populations annotated separately in the single-cell reference.

We mapped inferred stromal CXCL12 and ACTA2 expression in the Visium section (Fig. 4d,e). The Xenium analysis compared expression between the highest and lowest tertiles of neighborhood malignant cell fraction. The Visium analysis related expression to the fraction as a continuous variable, adjusting for the Cell2location stromal proportion and library size across modeled genes. FN1 and CXCL14 showed positive associations, whereas CXCL12, C3, MMP2, COL1A1, COL1A2, and COL3A1 showed negative associations (Fig. 4f). Across all three random seeds, partial Spearman correlations were close to zero for ACTA2 (0.008 to 0.014) and consistently negative for CXCL12 (−0.348 to −0.339). CXCL12 expression was lower in the high-fraction group in Xenium and showed a negative partial correlation with the neighborhood malignant cell fraction in Visium. ACTA2 expression was higher in the high-fraction group in Xenium and showed little association in the adjusted continuous analysis in Visium.

The two reference-defined populations, Macrophage 1 and Macrophage 2, differed in how their inferred gene expression was associated with the neighborhood malignant cell fraction. APOE expression showed a positive association in Macrophage 1 (partial Spearman correlations of 0.165–0.247 across seeds) and a weaker association in Macrophage 2 (0.034–0.077, Fig. 4g–i). For GPNMB, partial correlations were positive in Macrophage 1 (0.129–0.191) and near zero in Macrophage 2 (−0.066 to −0.025). For CD9, partial correlations were positive in Macrophage 2 (0.348–0.359) and near zero in Macrophage 1 (−0.056 to −0.050).

**Recovery of lesion-associated glial expression programs in multiple sclerosis**

The MS brain MERFISH benchmark tested whether SCATTER could recover glial expression programs associated with demyelinated lesions. Reactive astrocytes express intermediate filament and complement genes, and demyelinated regions contain fewer oligodendrocytes and less myelin[27,28]. Prior studies have also reported changes in oligodendrocyte subpopulations[29] and variation in reactive glial expression programs across tissue environments[30]. Using a cross-modality single-nucleus reference, we applied SCATTER to pseudo-spots from the MS MERFISH cohort[15]. The BANKSY[31] tissue-region labels distinguished demyelinated lesions, normal-appearing white matter, healthy white matter, and gray matter (Fig. 5a). Expression profiles derived from the constituent cells provided targets for evaluating recovery. The inferred glial programs were then compared across tissue regions and related to the original study's cell state annotations.

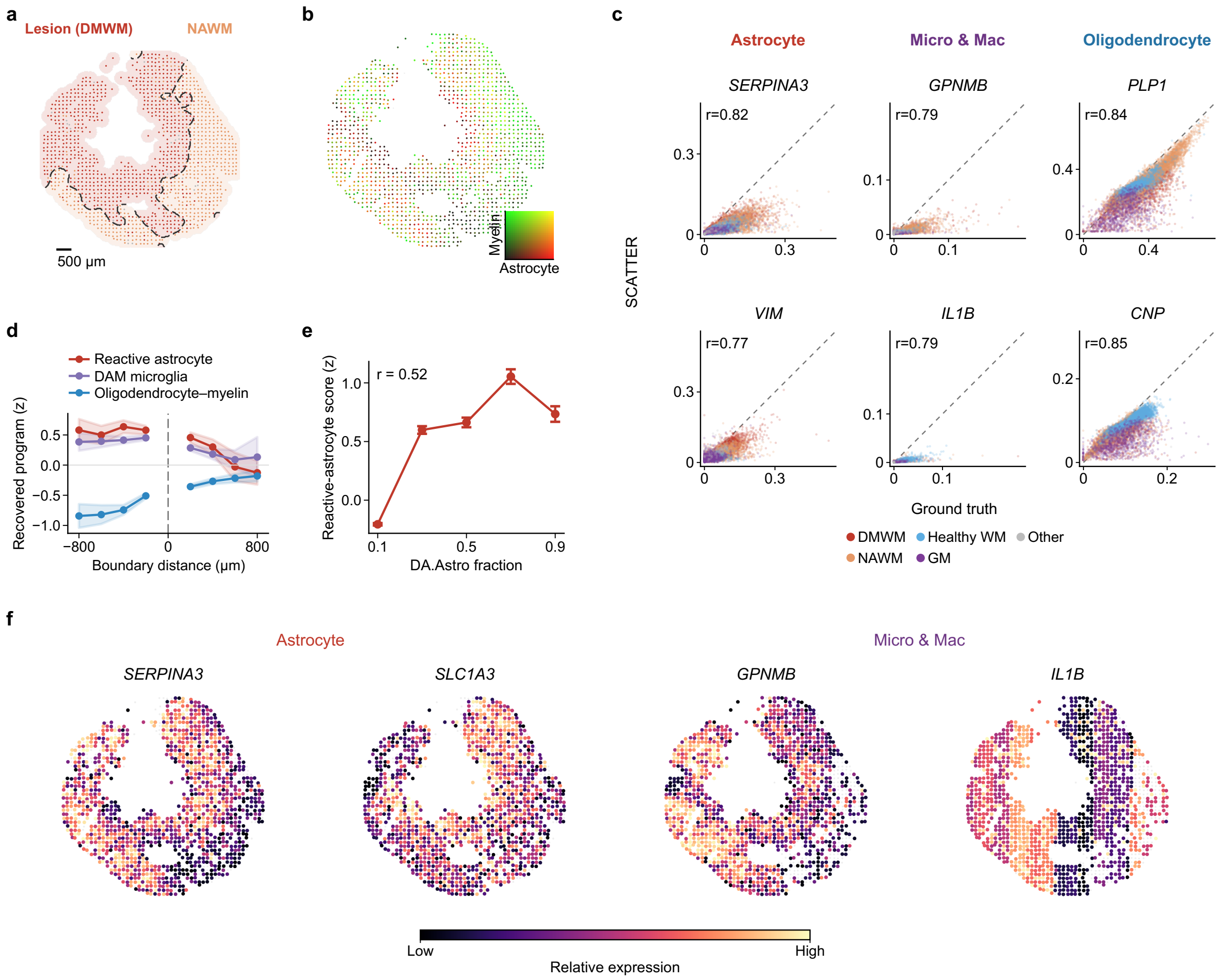

**Fig. 5. Recovery of glial expression near a multiple-sclerosis lesion boundary.** a, BANKSY tissue regions; the dashed line marks the lesion boundary. b, Reactive astrocyte (red) and oligodendrocyte–myelin (green) program scores shown as within-program percentile ranks. c, Inferred versus imaging-derived expression for six gene–cell type pairs. Points are eligible pseudo-spots, colors indicate tissue regions, the dashed line marks equality and r is the cohort-wide Pearson correlation. d, Reactive astrocyte, disease-associated microglial and oligodendrocyte–myelin program scores by signed distance from the lesion boundary; negative values are on the lesion side. e, Reactive astrocyte scores across eligible cohort pseudo-spots, grouped by the fraction of astrocytes annotated as disease-associated in the source study. f, Inferred spatial expression of SERPINA3 and SLC1A3 in astrocytes and GPNMB and IL1B in microglia–macrophages, ranked separately within each map. In d and e, lines show group means. Bands and error bars show mean ± 1.96 standard errors across spots and describe variation without accounting for spatial dependence. Maps show the same section. Scale bar in a, 500 μm. DMWM denotes demyelinated white matter, NAWM denotes normal-appearing white matter, WM denotes white matter and GM denotes gray matter. DAM denotes disease-associated microglia, "DA.Astro" denotes disease-associated astrocytes and "Micro & Mac" denotes microglia and macrophages.

Across eligible pseudo-spots, the reactive astrocyte program score based on inferred SERPINA3, VIM, C3, and CD44 expression correlated with the score calculated from the corresponding constituent cells (r = 0.74). At the single-gene level, correlations between inferred and constituent-cell-derived spatial expression ranged from r = 0.77 to 0.85 for the six illustrative glial gene–cell type pairs selected partly for high recovery (Fig. 5c and Supplementary Method 3). Supplementary Fig. 4a,b provides additional examples of single-gene expression recovery and compares regional mean inferred expression with the corresponding means derived from the constituent cells.

In a representative MS section, the recovered reactive astrocyte and oligodendrocyte–myelin expression programs occupied complementary regions (Fig. 5b). Figure 5f shows recovered SERPINA3, SLC1A3, GPNMB, and IL1B expression in that section. Across the full cohort, mean inferred expression in astrocytes was highest in demyelinated white matter for VIM and in gray matter for SLC1A3. In oligodendrocytes, CNP expression was highest in healthy white matter. The imaging-derived targets showed the same regional patterns (Supplementary Fig. 4b). Within the representative section, reactive astrocyte program scores were higher on the lesion side of the boundary than in the adjacent normal-appearing white matter (Fig. 5d). The oligodendrocyte–myelin program showed the reverse pattern (Fig. 5d). These spatial gradients resembled the glial expression patterns reported around chronic active MS lesions[15,27,32,33]. The section was selected for its balance of lesion and normal-appearing white matter among astrocyte-containing pseudo-spots. Supplementary Method 3 defines the cell-type proportion threshold for program scoring.

To relate the recovered astrocyte program to the cell states described in the original study, we compared the program score with the fraction of astrocytes in each pseudo-spot annotated as disease-associated. Across eligible pseudo-spots in the cohort, the two measures were positively correlated ($r = 0.52$). Figure 5e shows the grouped means. Previous reports have described disease-associated microglial expression programs near demyelinated tissue[27,34], so we also examined their spatial association with the recovered reactive astrocyte program. Across 9,643 eligible pseudo-spots pooled from the cohort, the inferred disease-associated microglial and reactive astrocyte program scores were positively correlated ($r = 0.37$, Supplementary Fig. 4d).

### Cell-type-specific expression at the colorectal cancer invasive front

The colorectal cancer analysis examined cell type–associated expression programs at the invasive tumor–stroma interface using Visium HD measurements from a human tissue section[4]. Measurements were aggregated into 16-μm spatial bins and analyzed with a matched Flex single-cell reference. The selected area included tumor nests, a 50-μm-wide invasion band, and the surrounding stroma (Fig. 6a). Previous colorectal cancer studies have described fibroblast–macrophage interactions near the invasive front[35] and the spatial organization of malignant, myeloid, and lymphoid populations[36]. We therefore examined the inferred expression programs of these populations at the tumor–stroma interface and related their spatial patterns to the original study's annotations and to candidate ligand–receptor pairs. Program scores were evaluated in bins where the inferred proportion of the corresponding cell type exceeded 0.05 (Supplementary Method 4).

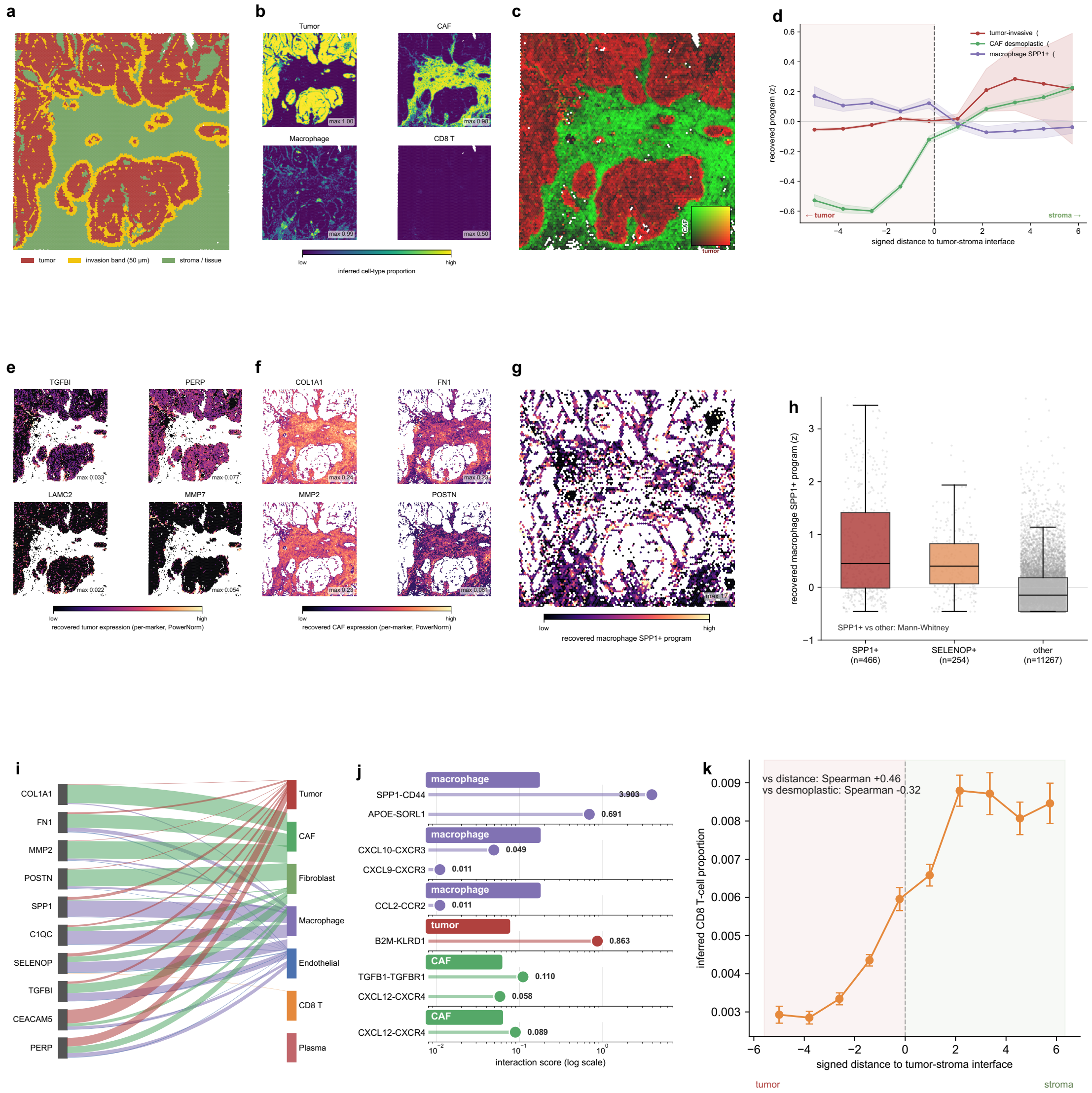

**Fig. 6. Cell-type-specific expression at the colorectal cancer invasive front in Visium HD.** a, Tumor region, 50-µm invasion band and outer tissue. b, Inferred proportions of four cell types on a common scale. c, Tumor invasion (red) and fibroblast desmoplastic (green) program scores as percentile ranks. d, Program scores by signed distance from the tumor–stroma interface; negative values are on the tumor side, and one coordinate unit is approximately 34.35 µm. e,f, Spatial maps of inferred tumor and fibroblast marker expression, each gene on its own scale. g, Inferred macrophage SPP1 program. h, The same program score by macrophage subtype as annotated in the source study (SPP1-positive, n = 466; SELENOP-positive, n = 254; other, n = 11,267). Points are spatial bins; boxes show medians and interquartile ranges, and whiskers extend to 1.5 times the interquartile range. The annotation reports a one-sided Mann–Whitney U comparison of SPP1-positive and other bins within this section. The test treats bins as independent and does not account for spatial dependence. i, Per-gene expression contributions across the seven displayed cell types in the invasion band, weighted by inferred proportions. j, Nine selected ligand–receptor pairs grouped by sender and receiver; scores are products of inferred ligand and receptor expression rates averaged across invasion-band spatial bins, multiplied by $10^6$ and displayed on a log scale. k, Inferred CD8 T-cell proportion by signed distance. The annotations give the Spearman correlations of CD8 T-cell proportion with signed distance and with desmoplastic program score. The latter uses bins eligible for fibroblast program scoring. For d and k, Spearman correlations use individual eligible spatial bins; curves show distance-group means, with bands and error bars showing mean ± 1.96 standard errors. CAF denotes cancer-associated fibroblasts.

At the tumor–stroma interface, the inferred tumor invasion program (including TGFBI, PERP, LAMC2, and MMP7) and the fibroblast-associated desmoplastic program (including COL1A1, FN1, MMP2, and POSTN) occupied adjacent regions along tumor nest boundaries (Fig. 6c,e,f). Across eligible spatial bins, signed distance from the tumor–stroma interface was positively correlated with the desmoplastic program score (Spearman ρ =

0.27). The tumor invasion program score showed little overall correlation with distance ($\rho = -0.06$, Fig. 6d). Correlations were computed across the individual spatial bins eligible for each program, whereas the curves in Fig. 6d show means within distance intervals. Collective invasion accompanied by partial epithelial-to-mesenchymal transition has also been observed in colorectal tumor organoids[37].

At the same invasive interface, the macrophage SPP1 program was concentrated along tumor nest edges (Fig. 6g). Its score was negatively correlated with signed distance across eligible spatial bins ($\rho = -0.28$). The median SPP1 program score was 0.45 in the 466 bins labeled Macrophage-SPP1+ and −0.15 in the 11,267 bins assigned to the "other macrophage" category in the original study. The rank-biserial correlation was 0.57 (one-sided Mann–Whitney U test, $P = 2.9 \times 10^{-97}$, Fig. 6h). These subtype annotations were withheld during model fitting. Among the seven cell types shown, macrophages had the largest estimated contributions to SPP1, C1QC, and SELENOP expression within the invasion band, with contributions weighted by cell-type proportion (Fig. 6i).

Nine selected candidate ligand–receptor pairs were ranked by their mean scores across the invasion band. In each spatial bin, the score was the product of the inferred ligand expression rate in the sending cell type and the inferred receptor expression rate in the receiving cell type. Among these nine pairs, SPP1 in macrophages and CD44 in tumor cells had the highest score, followed by tumor cell B2M–CD8 T-cell KLRD1, macrophage APOE–tumor cell SORL1, and fibroblast TGFB1–tumor cell TGFBR1 (Fig. 6j). The original study also highlighted the SPP1–CD44 pair[4]. A separate colorectal cancer study described interactions between SPP1-positive macrophages and FAP-positive fibroblasts[35].

We also examined how inferred CD8 T-cell proportions varied across the same tumor–stroma interface. Across individual bins, the inferred CD8 T-cell proportion increased with signed distance from tumor to stroma (Spearman $\rho = 0.46$) and was negatively associated with the desmoplastic program score in bins eligible for fibroblast program scoring ($\rho = -0.32$, Fig. 6k). The lower inferred CD8 T-cell proportion on the tumor side and the higher proportion on the stromal side are consistent with descriptions of immune exclusion[38].

## Discussion

SCATTER estimates cell-type proportions and spatially varying expression profiles from mixed spatial measurements. Its expression profiles describe how gene expression changes with location within each annotated cell population, extending analyses based on cell-type proportions. In each of the six pseudo-spot benchmarks, the mean within-cell-type spatial correlation was higher for SCATTER than for DestVI, PANDA, or STged. Its Pearson correlation for cell-type proportions was highest in three benchmarks and close to the best baseline in the other three. Agreement in spatial autocorrelation between predicted and target expression provided an additional assessment of the recovered spatial patterns.

In tissue sections, these profiles supported comparisons of expression patterns across cell populations in the same local environment. In the breast cancer Xenium pseudo-spots, comparison with targets derived from the constituent cells supported the recovery of gene-level associations with the tumor neighborhood and of expression-based regulon scores. In directly measured breast cancer Visium data, inferred APOE, GPNMB, and CD9 expression showed different associations with the tumor neighborhood in two reference-defined macrophage populations. In the representative MS section, the reactive astrocyte and oligodendrocyte–myelin programs showed opposing gradients across the lesion boundary. Across the cohort, the reactive astrocyte score was positively correlated with the fraction of astrocytes annotated as disease-associated in the original study ($r = 0.52$). At the colorectal cancer invasive front, macrophages had the largest estimated contributions to SPP1, C1QC, and SELENOP expression among the seven cell types shown. The macrophage SPP1 program was concentrated along tumor nest edges. Across these applications, the spot-level expression profiles provided information on within-cell-type variation in gene expression and expression programs beyond what cell-type proportions alone could provide.

Methods for expression recovery differ in how they represent variation within annotated populations. DestVI[10] learns continuous cell states from a single-cell reference, PANDA[11] uses mixtures of reference-derived archetypes, and STged[9] combines expression deconvolution with spatial and reference-based regularization. SpatialScope[39] infers expression and spatial organization at the level of individual cells, whereas SCATTER estimates a relative expression profile for each cell type at each spot. SCATTER's spot-specific expression profiles can be used to examine associations with spatial covariates, whereas C-SIDE[40] directly estimates cell-type-specific differential expression associated with prespecified covariates.

Deconvolving expression from mixed counts requires assumptions about the contributing cell populations. For each gene at a given spot, the data contain one observed count and several unknown cell-type contributions, so cell-type-specific expression profiles are not identifiable without additional information. Residual backfitting depends on the fitted proportions and locally estimated profiles. Errors in the proportions can affect the residuals attributed to multiple cell types. Cell2location's cell abundance estimates are normalized to obtain the proportion prior. SCATTER uses fitted cell-type proportions as weights in the transcript mixture during neural network fitting. Differences in RNA content per cell can affect how closely those proportions represent each type's contribution to the mixed transcript counts. The final expression profiles are normalized across measured genes within each cell type and therefore represent relative expression over the modeled gene panel. Regulon scores computed from inferred profiles summarize relative expression of measured target genes. Candidate ligand–receptor pairs are ranked using products of inferred relative expression in the corresponding cell types.

The six primary pseudo-spot benchmarks assessed expression recovery using mixtures with known cell-type compositions. The five MERFISH and Xenium settings used the same single-cell measurements for mixed counts and expression targets. Stereo-seq used disjoint transcript pools from the same constituent cells. The breast cancer benchmark used a reference with no overlap with the imaged cells and was restricted to the 313 genes shared by the Xenium data and the scFFPE-seq reference. The directly measured breast cancer Visium and colorectal cancer Visium HD data had no matched single-cell-resolution expression targets. The lung Visium HD evaluation used Xenium measurements from the same or adjacent sections as a cross-platform reference (Table 2 and Supplementary Table 9). The sci-Space comparison tested recovery from a separate set of constructed pseudo-spots (Supplementary Table 10). The primary benchmark analyses used random seed 0. The directly measured breast cancer Visium analysis was repeated on the same section using three seeds. Further evaluation could include additional tissues and platforms, perturbations of the reference data, and quantification of uncertainty. Within the scope of this evaluation, improved expression recovery in paired benchmark comparisons and the analyses of directly measured tissue data support the use of cell-type-specific profiles to study transcriptional variation across tissue microenvironments.

# Methods

## Datasets and preprocessing

We constructed six pseudo-spot benchmarks from spatial transcriptomics datasets that provide gene expression measurements, spatial coordinates, and cell-type annotations at single-cell resolution. Cell coordinates determined assignment to non-overlapping square bins. For hypothalamic MERFISH and Xenium, the coordinates were cell centroids. Bins with fewer than the dataset-specific minimum number of cells were discarded. For each gene, pseudo-spot inputs summed all retained cellular counts in the five MERFISH and Xenium settings and only the input-pool counts in Stereo-seq. In the five MERFISH and Xenium settings, pseudo-spot coordinates were the mean coordinates of their constituent cells. For MS, we calculated the coordinates separately within each sample and then separated the samples in a common coordinate system. The ground-truth proportion for a cell type was the fraction of all cells in the pseudo-spot assigned to that type. Expression targets were calculated from cell profiles normalized by library size. For each pseudo-spot and cell type, the average of those profiles was rescaled to sum to 1 across genes. The target tensor was withheld during model fitting. The mixed counts and expression targets were derived from the same constituent cells. For Stereo-seq, expression targets used transcripts withheld from the mixed counts. Predicted and target expression were evaluated as gene expression rates within each spot–cell type combination. When cells have different library sizes, mixing target profiles in proportion to cell counts may fail to reconstruct the normalized mixed counts exactly, even though all cells contribute equally when computing the targets.

The whole-brain benchmark used coronal section 30 from the Allen Brain Cell Atlas MERFISH dataset[12]. Eligibility required coordinates, subclass labels, and a subclass represented by at least 100 cells in the spatial data. After filtering, we assigned the remaining cells to square bins with a side length of 0.15 in the original x and y coordinate units and retained bins containing at least five cells. A subclass-stratified subsample of 50,000 cells from the external Allen Brain Cell Atlas WMB-10Xv3 dataset provided the reference, with the analysis restricted to shared cell subclasses and genes. The hypothalamic benchmark used MERFISH data from the medial preoptic area[13] and the Moffitt single-cell RNA-seq reference distributed with STged for the merged cell categories. Centroid_X and Centroid_Y determined assignment to 100-μm square bins. Bins containing at least two cells were retained. The breast cancer benchmark used a 10x Xenium sample and cell labels obtained by supervised annotation[14]. The analysis included cell types and genes shared with an independent scFFPE-seq reference with no overlapping cells, yielding a panel of 313 genes. Xenium cells were assigned to 100-μm bins by x_centroid and y_centroid. Bins containing at least five cells were retained. The MS benchmarks used 17 samples from the GSE284005 MERSCOPE MERFISH cohort[15] and their majorCluster_final annotations. Within each sample, cell coordinates determined assignment to 150-μm square bins. Bins containing at least ten cells were retained. An offset along the x axis left a gap of ten bin widths between samples. The cross-modality benchmark used the 477 genes shared by the 500-gene MERFISH panel and the GSE301908 single-nucleus reference. For the same-modality benchmark, a split within each cell type (random seed 0) assigned 25% of the cells to the MERFISH reference and made the remaining 75% available for pseudo-spot construction. The reference was capped at 8,000 cells per cell type, and the two subsets did not overlap.

The Stereo-seq benchmark used the adult mouse hemibrain cell-bin data from the MOSTA resource[16] and the source cell annotations, grouped into seven lineages: glutamatergic neurons, GABAergic neurons, astrocytes, oligodendrocytes, oligodendrocyte precursor cells, immune cells and vascular cells. Cells annotated as unknown, erythrocytes, ependymal cells or dopaminergic neurons were excluded. The reference was drawn from the 50,000-cell Allen Brain Cell Atlas WMB-10Xv3 subsample used for the whole-brain benchmark. We grouped the reference subclasses into the same seven lineages and retained up to 2,000 cells per lineage, giving 9,125 reference cells. The analysis used 3,000 highly variable genes among those shared with the reference. The integer transcript counts of each cell were split into two disjoint pools by binomial sampling with probability 0.5. One pool supplied the mixed counts and the other supplied the expression targets. Cells were assigned to 100-μm square bins using the source coordinates (0.5 μm per unit), bins with at least three cells were retained and pseudo-spot coordinates were the bin positions. The normalized targets were then restricted to the final panel of 2,935 genes with nonzero input and reference counts. The benchmark contained 2,209 pseudo-spots.

Marker-gene concordance was evaluated in two directly measured sequencing-based tumor datasets from stage III melanoma[17] and cutaneous squamous cell carcinoma[18]. Neither dataset provided single-cell-resolution expression targets. The breast cancer Visium CytAssist analysis in Fig. 4 used a section and an scFFPE-seq reference from the same FFPE tissue block[14]. The colorectal cancer Visium HD analysis in Fig. 6 used section data cropped to the invasive front and aggregated into 16-μm bins[4]. Model inputs for each dataset were a spot × gene count matrix, spatial coordinates, and annotated single-cell reference data, with gene expression restricted to the shared panel. Supplementary Table 7 lists the dataset dimensions and the numbers of cells used to construct the pseudo-spots.

## SCATTER

A slide contains $S$ spots, $G$ genes and $K$ annotated cell types. The observed count matrix is $y \in \mathbb{N}^{S\times G}$, with spot library size $\ell_s = \sum_g y_{s,g}$ and spatial coordinates $r_s \in \mathbb{R}^2$. The single-cell reference contains $N$ cells with annotated cell types and rate-normalized expression $X \in [0,1]^{N\times G}$, where each row $x_c$ satisfies $\sum_g x_{c,g} = 1$. SCATTER estimates cell-type proportions $\alpha_{s,\cdot} \in \Delta^{K-1}$ and expression profiles $E_{s,k,\cdot} \in \Delta^{G-1}$ for each spot and cell type, where $\Delta^{d-1}$ denotes the simplex of non-negative vectors of length $d$ that sum to one. The model uses two sequential fitting stages. The first estimates cell-type composition through softmax reference coupling, and the second fits expression profiles through second-order Taylor reference coupling while holding the first-stage composition estimates fixed. Superscripts $(1)$ and $(2)$ denote quantities and parameters belonging to the first and second fitting stages, respectively. The architecture defined below is instantiated separately in each stage, with independently fitted parameters. Equations without stage superscripts specify the shared architecture or loss definitions and apply to the corresponding stage-specific quantities.

The spot encoder uses library-size-normalized, log-transformed counts,

$$\tilde{y}_{s,g} = \log\left(1 + 10^4 \frac{y_{s,g}}{\ell_s}\right).$$

Coordinates are independently rescaled to $\bar{r}_s \in [0,1]^2$ along each axis and mapped to fixed Gaussian random Fourier features,

$$p_s = \left[\sin\left(2\pi \bar{r}_s^{\top}\omega\right), \cos\left(2\pi \bar{r}_s^{\top}\omega\right)\right], \qquad \omega \in \mathbb{R}^{2\times M}.$$

We use $M = 32$ frequencies, with entries of $\omega$ independently sampled from $\mathcal{N}(0, 5^2)$ and held fixed during training [41,42]. An input multilayer perceptron (MLP), $f_{\text{in}}$, combines the transformed count profile and positional features into a hidden state of dimension $d_h$,

$$h_s^{(0)} = f_{\text{in}}(\tilde{y}_s, p_s).$$

A graph-transformer block applies scaled dot-product attention over $\mathcal{N}(s) = \{s\} \cup \text{NN}_8(r_s)$, where $\text{NN}_8(r_s)$ denotes the eight nearest other spots according to the prepared coordinates $r_s$. Neighbor indices are computed before axis-wise coordinate rescaling. The block uses four attention heads, post-normalization residual connections and one message-passing layer, yielding the contextual state $\bar{h}_s \in \mathbb{R}^{d_h}$. During the first fitting stage, an affine composition head $f_\alpha$ maps this state to cell-type proportions,

$$\alpha_{s,\cdot} = \text{softmax}\left(f_\alpha(\bar{h}_s)\right).$$

The output MLP $f_{\text{out}}$ maps the contextual state to a matching vector of dimension $d_g$, which is divided by its Euclidean norm,

$$\hat{h}_s = \text{L2norm}\left(f_{\text{out}}(\bar{h}_s)\right) \in \mathbb{R}^{d_g}.$$

The reference MLP $f_{\text{ref}}$ maps each log-transformed reference profile to an embedding in the same space, followed by the same L2 normalization,

$$u_c = \text{L2norm}\left(f_{\text{ref}}\left(\log(1 + 10^4 x_c)\right)\right).$$

For each cell type $k$, an affine projection $A_k$ maps the contextual spot state to a latent code of dimension $d_z$,

$$z_{s,k} = A_k(\overline{h}_s) \in \mathbb{R}^{d_z}.$$

A learned matrix $B_k$ maps this code to an offset $\delta_{s,k}$ in the matching space. A learned type vector $v_k$ and this offset are added to the shared spot embedding to form the type-specific query,

$$\delta_{s,k} = B_k z_{s,k}, \qquad \hat{h}_{s,k} = \mathrm{L2norm}(\hat{h}_s + v_k + \delta_{s,k}).$$

The weights of $A_k$ are initialized near zero, its bias is initialized to zero, and $B_k$ is initialized at a scale of 0.01. Let $\mathcal{C}_k$ be the reference cells annotated as type $k$. In the first fitting stage, a learned temperature $\tau > 0$ scales the cosine similarity between each type-specific query and reference embedding. Softmax normalization over cells of the same annotated type gives the coupling weights,

$$q_{s,c|k}^{(1)} = \mathrm{softmax}_{c \in \mathcal{C}_k}\left(\frac{\left\langle \hat{h}_{s,k}^{(1)}, u_c^{(1)} \right\rangle}{\tau^{(1)}}\right).$$

The first-stage reference-derived expression profile is the weighted sum of normalized reference-cell profiles,

$$E_{s,k}^{\mathrm{hull},(1)} = \sum_{c \in \mathcal{C}_k} q_{s,c|k}^{(1)}\, x_{c,\cdot}.$$

A learned gene-specific calibration factor $e^{\gamma_g^{(1)}}$ scales the reconstructed counts. The predicted mean for gene $g$ in spot $s$ during the first fitting stage is

$$\mu_{s,g}^{(1)} = \ell_s e^{\gamma_g^{(1)}} \sum_k \alpha_{s,k}^{(1)}\, E_{s,k,g}^{\mathrm{hull},(1)}.$$

The count likelihood uses conditionally independent negative-binomial draws $y_{s,g} \mid \mu_{s,g}, \rho_g \sim \mathrm{NB}(\mu_{s,g}, \rho_g)$, parameterized by the predicted mean $\mu_{s,g}$ and learned per-gene dispersion $\rho_g > 0$. Omitting terms independent of the model parameters, the mean negative log-likelihood is

$$\mathcal{L}_{\mathrm{NB}} = -\frac{1}{SG}\sum_{s,g}\left[\log\Gamma(y_{s,g} + \rho_g) - \log\Gamma(\rho_g) + \rho_g \log\frac{\rho_g}{\rho_g + \mu_{s,g}} + y_{s,g}\log\frac{\mu_{s,g}}{\rho_g + \mu_{s,g}}\right].$$

The log-space reconstruction loss compares observed and predicted counts,

$$\mathcal{L}_{\mathrm{rec}} = \frac{1}{SG}\sum_{s,g}\left[\log(1 + y_{s,g}) - \log(1 + \mu_{s,g})\right]^2.$$

Before SCATTER fitting, Cell2location is fitted to the same prepared slide to obtain a composition prior $\beta_s \in \Delta^{K-1}$ for each spot, normalized over cell types. During the first fitting stage, the composition loss is

$$\mathcal{L}_{\mathrm{comp}} = \frac{1}{S}\sum_s \mathrm{KL}\left(\beta_s \parallel \alpha_s^{(1)}\right), \qquad \mathrm{KL}\left(\beta_s \parallel \alpha_s^{(1)}\right) = \sum_k \beta_{s,k}\log\left(\frac{\beta_{s,k}}{\alpha_{s,k}^{(1)} + \epsilon_0}\right).$$

The small constant $\epsilon_0$ is $10^{-9}$ in the denominator. Terms with zero prior weight contribute zero to the loss. At optimization step $t$, the weight $w(t)$ decreases linearly from $w_0$ to $w_{\mathrm{floor}}$ over $T$ steps,

$$w(t) = w_{\mathrm{floor}} + (w_0 - w_{\mathrm{floor}})\max\left(0, 1 - \frac{t}{T}\right).$$

For the encoder-stability loss, independent coordinate noise $\epsilon_s \sim \mathcal{N}(0, \sigma^2 I_2)$ is drawn at each training step, where $\sigma$ is the perturbation standard deviation. The count input and neighbor indices remain fixed while the embeddings from the original and perturbed coordinates are compared,

$$\mathcal{L}_{\text{enc}} = \frac{1}{S}\sum_{s} \left\|\hat{h}_s(\bar{r}_s) - \hat{h}_s(\bar{r}_s + \epsilon_s)\right\|_2^2.$$

Latent codes are regularized over the same spatial graph. With edge set $\mathcal{E} = \{(s,j): j \in \mathcal{N}(s), j \neq s\}$, the loss is

$$\mathcal{L}_z = \frac{1}{|\mathcal{E}|K}\sum_{k}\sum_{(s,j)\in\mathcal{E}} \left\|z_{s,k} - z_{j,k}\right\|_2^2.$$

With fixed weights $\lambda_{\text{rec}}$, $\lambda_{\text{enc}}$ and $\lambda_z$, the first-stage training objective is

$$\mathcal{L}^{(1)} = \mathcal{L}_{\text{NB}}^{(1)} + \lambda_{\text{rec}}\mathcal{L}_{\text{rec}}^{(1)} + w(t)\mathcal{L}_{\text{comp}}^{(1)} + \lambda_{\text{enc}}\mathcal{L}_{\text{enc}}^{(1)} + \lambda_z\mathcal{L}_z^{(1)}.$$

All first-stage model parameters are jointly optimized using full-batch Adam for 3,000 steps. The learning rate is 0.002 for the first 2,400 steps and 0.0005 for the remaining 600 steps, with gradient norms clipped to 5.0. After fitting, the final composition estimates are saved as

$$\hat{\alpha}_{s,k} = \alpha_{s,k}^{(1)}.$$

The second fitting stage uses a separately initialized model with the same architecture, random seed and parameter initialization procedure as the first stage. Its trainable parameters are optimized independently of the fitted first-stage parameters. The saved composition estimates $\hat{\alpha}_{s,k}$ are held fixed, and the second-stage composition head is inactive and frozen. The spot encoder, reference encoder, type-specific query construction, latent projections and gene calibration retain the definitions given above, with independently optimized second-stage parameters.

In this stage, reference coupling is computed using a second-order Taylor polynomial of the temperature-scaled cosine similarity. For each spot $s$, reference cell $c \in \mathcal{C}_k$ and annotated cell type $k$, define

$$a_{s,c|k}^{(2)} = \frac{\left\langle \hat{h}_{s,k}^{(2)}, u_c^{(2)} \right\rangle}{\tau^{(2)}}, \qquad \tau^{(2)} > 0.$$

The second-stage coupling weights are

$$q_{s,c|k}^{(2)} = \frac{1 + a_{s,c|k}^{(2)} + \frac{1}{2}\left(a_{s,c|k}^{(2)}\right)^2}{\sum_{c'\in\mathcal{C}_k}\left[1 + a_{s,c'|k}^{(2)} + \frac{1}{2}\left(a_{s,c'|k}^{(2)}\right)^2\right]}.$$

These weights are normalized over the reference cells annotated as type $k$. The resulting reference-derived expression profile is

$$E_{s,k}^{\text{hull},(2)} = \sum_{c\in\mathcal{C}_k} q_{s,c|k}^{(2)}\, x_{c,\cdot}.$$

At a given spot, consider two reference cells of the same type with temperature-scaled matching scores $a_i$ and $a_j$. Their Taylor weights $q_i$ and $q_j$ share a normalization factor, so $\log(q_i/q_j) = \log f(a_i) - \log f(a_j)$, where $f(a) = 1 + a + a^2/2$. The logarithm of the Taylor polynomial satisfies

$$\left|\frac{\mathrm{d}}{\mathrm{d}a}\log f(a)\right| = \frac{2|a+1|}{1 + (a+1)^2} \leq 1.$$

Equality in the derivative bound occurs only at $a = 0$ and $a = -2$. Integrating between $a_i$ and $a_j$ therefore gives $|\log(q_i/q_j)| < |a_i - a_j|$ for distinct scores. For the same scores, softmax gives a log weight ratio of $a_i - a_j$.

With the first-stage composition estimates fixed, the second-stage predicted count mean is

$$\mu_{s,g}^{(2)} = \ell_s e^{\gamma_g^{(2)}} \sum_k \hat{\alpha}_{s,k} \, E_{s,k,g}^{\mathrm{hull},(2)}.$$

The second-stage negative-binomial and log-space reconstruction losses use $\mu_{s,g}^{(2)}$, with independently learned gene-specific calibration parameters $\gamma_g^{(2)}$ and dispersions $\rho_g^{(2)}$. The encoder-stability and latent-code smoothness losses are calculated using the second-stage embeddings and latent codes, respectively. Because the composition estimates are fixed, the composition loss does not contribute to second-stage optimization. The second-stage objective is

$$\mathcal{L}^{(2)} = \mathcal{L}_{\mathrm{NB}}^{(2)} + \lambda_{\mathrm{rec}} \mathcal{L}_{\mathrm{rec}}^{(2)} + \lambda_{\mathrm{enc}} \mathcal{L}_{\mathrm{enc}}^{(2)} + \lambda_z \mathcal{L}_z^{(2)}.$$

All active second-stage parameters, including the spot and reference encoders, matching parameters, temperature, gene calibration and dispersion, are jointly optimized using an independent full-batch Adam optimizer. Training again uses 3,000 steps, with a learning rate of 0.002 for the first 2,400 steps and 0.0005 thereafter, and gradient norms clipped to 5.0.

After the second fitting stage, the gene-specific calibration factors are applied to each reference-derived expression profile, which is then normalized over genes,

$$E_{s,k,g}^{\mathrm{cal}} = \frac{e^{\gamma_g^{(2)}} E_{s,k,g}^{\mathrm{hull},(2)}}{\sum_{g'} e^{\gamma_{g'}^{(2)}} E_{s,k,g'}^{\mathrm{hull},(2)}}.$$

Residual backfitting starts from $E_{s,k}^{(0)} = E_{s,k}^{\mathrm{cal}}$ and uses the observed spot rate $\bar{y}_s = y_s/\ell_s$. Before backfitting, each fitted proportion is bounded below by $10^{-6}$ and the proportions are renormalized within each spot. Throughout the sweeps, these adjusted composition estimates $\hat{\alpha}_{s,k}$ remain fixed. During each sweep $m$, $E_{s,k}$ denotes the current expression profile. Each update subtracts the current estimated contributions of the other cell types,

$$\mathrm{res}_{s,k} = \bar{y}_s - \sum_{k' \neq k} \hat{\alpha}_{s,k'} \, E_{s,k'}.$$

The residual is rectified and normalized across genes, then blended with the current profile using coefficient $b \in (0,1]$,

$$E_{s,k}^{\mathrm{res}} = \mathrm{normalize}\left(\mathrm{relu}\left(\frac{\mathrm{res}_{s,k}}{\hat{\alpha}_{s,k}}\right)\right), \qquad E_{s,k} \leftarrow (1-b)E_{s,k} + bE_{s,k}^{\mathrm{res}}.$$

Here relu sets negative entries to zero, and normalize divides a non-negative vector by its gene sum; a zero-sum vector remains zero. Each sweep updates cell types in the column order of the composition matrix, using the most recently revised profiles in subsequent residual calculations. We set $b = 0.15$. Let $E^{(m)}$ denote all profiles after completing sweep $m$. Backfitting stops when

$$\frac{\|E^{(m)} - E^{(m-1)}\|_F}{\|E^{(m-1)}\|_F} < 0.01,$$

with a maximum of 16 sweeps, where $\|\cdot\|_F$ is the Frobenius norm over the full profile tensor. The final profiles are normalized over genes.

SCATTER returns the first-stage cell-type proportions $\hat{\alpha}_{s,k}$ and the expression profiles obtained from second-stage fitting and residual backfitting. Default architecture, training and backfitting settings are listed in Supplementary Table 8.

### Baseline methods

We selected baseline methods according to the output being evaluated (Supplementary Note 1). For cell-type proportions, we compared SCATTER with NNLS, STINR[6], Cell2location[5], DestVI[10], PANDA[11], and STged[9]. For the proportion benchmark, we represented each method's estimates as a spot × cell type matrix of proportions. NNLS fits reference-derived expression signatures by nonnegative least squares, whereas STINR estimates spatially varying proportions using a fixed expression basis for each cell type. Cell2location serves as a probabilistic deconvolution baseline and provides the proportion prior used to train SCATTER. DestVI, PANDA, and STged also return expression profiles for each cell type at each spatial location, which allowed us to include them in the comparisons of expression recovery and within-cell-type spatial autocorrelation. For NNLS, STINR, and Cell2location, we evaluated cell-type proportions only. NNLS and STINR use fixed cell-type expression profiles, and Cell2location does not return a spot-by-cell-type expression tensor. All methods used the shared gene panel. Supplementary Table 6 describes the implementation settings and any departures from published workflows. Supplementary Notes 2 and 3 and Supplementary Methods 1–4 detail the evaluation and downstream analyses.

## Data availability

All source datasets are publicly available. These include whole-brain MERFISH data from the Allen Brain Cell Atlas[12], hypothalamic preoptic area MERFISH data[13] (Dryad, doi:10.5061/dryad.8t8s248), human breast Xenium and directly measured Visium CytAssist data with their scFFPE-seq reference[14] (GEO GSE243280, comprising GSE243168 for Xenium and GSE243275 for single-cell RNA-seq and Visium), MS brain MERFISH data (GEO GSE284005[15]) and the matched single-nucleus reference (GEO GSE301908), adult mouse hemibrain Stereo-seq data[16] (MOSTA dataset STDS0000058, Mouse_brain_cell_bin.h5ad), human lung cancer Visium HD and Xenium data (10x Genomics datasets Visium_HD_Human_Lung_Cancer_post_Xenium_v1_Experiment1, Visium_HD_Human_Lung_Cancer_HD_Only_Experiment1, Visium_HD_Human_Lung_Cancer_HD_Only_Experiment2, Xenium_V1_Human_Lung_Cancer_FFPE and Xenium_Prime_Human_Lung_Cancer_FFPE) with the lung adenocarcinoma single-cell reference[43] (GEO GSE131907), sci-Space mouse embryo data[44] (GEO GSE166692), colorectal cancer Visium HD data and the matched Flex single-cell reference (GEO GSE280318[4]), stage III melanoma data[17] (PANDA release, Zenodo 13323363), and cutaneous squamous cell carcinoma data[18] (STged release, Zenodo 14379873, GEO GSE144240). The processed datasets used in this study will be deposited in Zenodo upon publication.

## Code availability

Code for reconstructing the processed datasets and reproducing the analyses and figures will be made publicly available upon publication through the project repository (https://github.com/learnslowly/scatter). We will release the code and a software environment with pinned package versions in a tagged repository release and archive that release on Zenodo. SCATTER was trained on a single GPU. Cell2location and DestVI also used GPUs. NNLS, STINR, PANDA, STged, and the downstream analyses used CPU resources on a Slurm cluster.

## Acknowledgments

Part of this work used high-performance computing resources provided by the Louisiana Optical Network Infrastructure (http://www.loni.org).

This work was also supported in part by high-performance computing resources and services provided by Tulane University Information Technology in New Orleans, Louisiana, USA.

## Author contributions

Lei Huang conceived the method, performed all experiments and analyses, and drafted the manuscript. Yun Gong, Weiqiang Lin, and Lindong Jiang revised the manuscript and provided input on the experiments. Hui Shen revised the manuscript. Hong-Wen Deng and Chaoyang Zhang supervised the study.

## Competing interests

The authors declare no competing interests.

## Ethics statement

This study analyzed publicly available research and technical datasets. The original publications and data providers document the provenance of the tissue samples. We collected no new human or animal data for this study.

## Additional information

Supplementary information accompanies this article.

Correspondence and requests for materials should be addressed to the corresponding authors.

## References

1. Rao, A., Barkley, D., França, G. S. & Yanai, I. Exploring tissue architecture using spatial transcriptomics. *Nature* **596**, 211–220 (2021).

2. You, Y. *et al.* Systematic comparison of sequencing-based spatial transcriptomic methods. *Nat. Methods* **21**, 1743–1754 (2024).

3. Ståhl, P. L. *et al.* Visualization and analysis of gene expression in tissue sections by spatial transcriptomics. *Science* **353**, 78–82 (2016).

4. Oliveira, M. F. D. *et al.* High-definition spatial transcriptomic profiling of immune cell populations in colorectal cancer. *Nat. Genet.* **57**, 1512–1523 (2025).

5. Kleshchevnikov, V. *et al.* Cell2location maps fine-grained cell types in spatial transcriptomics. *Nat. Biotechnol.* **40**, 661–671 (2022).

6. Luo, Y., Zhao, X., Ye, K. & Meng, D. STINR: Deciphering Spatial Transcriptomics via Implicit Neural Representation. in *2025 IEEE/CVF Conference on Computer Vision and Pattern Recognition (CVPR)* 25930–25939 (IEEE, Nashville, TN, USA, 2025). doi:10.1109/CVPR52734.2025.02415.

7. Cable, D. M. *et al.* Robust decomposition of cell type mixtures in spatial transcriptomics. *Nat. Biotechnol.* **40**, 517–526 (2022).

8. Ma, Y. & Zhou, X. Spatially informed cell-type deconvolution for spatial transcriptomics. *Nat. Biotechnol.* **40**, 1349–1359 (2022).

9. Tu, J.-J., Yan, H., Zhang, X.-F. & Lin, Z. Precise gene expression deconvolution in spatial transcriptomics with STged. *Nucleic Acids Res.* **53**, gkaf087 (2025).

10. Lopez, R. *et al.* DestVI identifies continuums of cell types in spatial transcriptomics data. *Nat. Biotechnol.* **40**, 1360–1369 (2022).

11. Wang, M.-G., Chen, L. & Zhang, X.-F. Dual decoding of cell types and gene expression in spatial transcriptomics with PANDA. *Nucleic Acids Res.* **52**, 12173–12190 (2024).

12. Yao, Z. *et al.* A high-resolution transcriptomic and spatial atlas of cell types in the whole mouse brain. *Nature* **624**, 317–332 (2023).

13. Moffitt, J. R. *et al.* Molecular, spatial, and functional single-cell profiling of the hypothalamic preoptic region. *Science* **362**, eaau5324 (2018).

14. Janesick, A. *et al.* High resolution mapping of the tumor microenvironment using integrated single-cell, spatial and in situ analysis. *Nat. Commun.* **14**, 8353 (2023).

15. Feng, R. *et al.* Single-cell spatial transcriptomic profiling defines a pathogenic inflammatory niche in chronic active multiple sclerosis lesions. *Immunity* **58**, 2989-3005.e10 (2025).

16. Chen, A. *et al.* Spatiotemporal transcriptomic atlas of mouse organogenesis using DNA nanoball-patterned arrays. *Cell* **185**, 1777-1792.e21 (2022).

17. Thrane, K., Eriksson, H., Maaskola, J., Hansson, J. & Lundeberg, J. Spatially Resolved Transcriptomics Enables Dissection of Genetic Heterogeneity in Stage III Cutaneous Malignant Melanoma. *Cancer Res.* **78**, 5970–5979 (2018).

18. Ji, A. L. *et al.* Multimodal Analysis of Composition and Spatial Architecture in Human Squamous Cell Carcinoma. *Cell* **182**, 497-514.e22 (2020).

19. Carmeliet, P. & Jain, R. K. Molecular mechanisms and clinical applications of angiogenesis. *Nature* **473**, 298–307 (2011).

20. Wang, X. *et al.* Spatially resolved atlas of breast cancer uncovers intercellular machinery of venular niche governing lymphocyte extravasation. *Nat. Commun.* **16**, 3348 (2025).

21. Sirka, O. K., Shamir, E. R. & Ewald, A. J. Myoepithelial cells are a dynamic barrier to epithelial dissemination. *J. Cell Biol.* **217**, 3368–3381 (2018).

22. Risom, T. *et al.* Transition to invasive breast cancer is associated with progressive changes in the structure and composition of tumor stroma. *Cell* **185**, 299-310.e18 (2022).

23. DeNardo, D. G. & Ruffell, B. Macrophages as regulators of tumour immunity and immunotherapy. *Nat. Rev. Immunol.* **19**, 369–382 (2019).

24. Orimo, A. *et al.* Stromal Fibroblasts Present in Invasive Human Breast Carcinomas Promote Tumor Growth and Angiogenesis through Elevated SDF-1/CXCL12 Secretion. *Cell* **121**, 335–348 (2005).

25. Wu, S. Z. *et al.* A single-cell and spatially resolved atlas of human breast cancers. *Nat. Genet.* **53**, 1334–1347 (2021).

26. Garcia-Alonso, L., Holland, C. H., Ibrahim, M. M., Turei, D. & Saez-Rodriguez, J. Benchmark and integration of resources for the estimation of human transcription factor activities. *Genome Res.* **29**, 1363–1375 (2019).

27. Absinta, M. *et al.* A lymphocyte–microglia–astrocyte axis in chronic active multiple sclerosis. *Nature* **597**, 709–714 (2021).

28. Liddelow, S. A. *et al.* Neurotoxic reactive astrocytes are induced by activated microglia. *Nature* **541**, 481–487 (2017).

29. Jäkel, S. *et al.* Altered human oligodendrocyte heterogeneity in multiple sclerosis. *Nature* **566**, 543–547 (2019).

30. Escartin, C. *et al.* Reactive astrocyte nomenclature, definitions, and future directions. *Nat. Neurosci.* **24**, 312–325 (2021).

31. Singhal, V. *et al.* BANKSY unifies cell typing and tissue domain segmentation for scalable spatial omics data analysis. *Nat. Genet.* **56**, 431–441 (2024).

32. Kukanja, P. *et al.* Cellular architecture of evolving neuroinflammatory lesions and multiple sclerosis pathology. *Cell* **187**, 1990-2009.e19 (2024).

33. Franklin, R. J. M. & ffrench-Constant, C. Remyelination in the CNS: from biology to therapy. *Nat. Rev. Neurosci.* **9**, 839–855 (2008).

34. Lerma-Martin, C. *et al.* Cell type mapping reveals tissue niches and interactions in subcortical multiple sclerosis lesions. *Nat. Neurosci.* **27**, 2354–2365 (2024).

35. Qi, J. *et al.* Single-cell and spatial analysis reveal interaction of FAP+ fibroblasts and SPP1+ macrophages in colorectal cancer. *Nat. Commun.* **13**, 1742 (2022).

36. Pelka, K. *et al.* Spatially organized multicellular immune hubs in human colorectal cancer. *Cell* **184**, 4734-4752.e20 (2021).

37. Flum, M. *et al.* Canonical TGFβ signaling induces collective invasion in colorectal carcinogenesis through a Snail1- and Zeb1-independent partial EMT. *Oncogene* **41**, 1492–1506 (2022).

38. Feng, Y. *et al.* Spatially organized tumor-stroma boundary determines the efficacy of immunotherapy in colorectal cancer patients. *Nat. Commun.* **15**, 10259 (2024).

39. Wan, X. *et al.* Integrating spatial and single-cell transcriptomics data using deep generative models with SpatialScope. *Nat. Commun.* **14**, 7848 (2023).

40. Cable, D. M. *et al.* Cell type-specific inference of differential expression in spatial transcriptomics. *Nat. Methods* **19**, 1076–1087 (2022).

41. Rahimi, A. & Recht, B. Random Features for Large-Scale Kernel Machines. in *Advances in Neural Information Processing Systems* (eds Platt, J., Koller, D., Singer, Y. & Roweis, S.) vol. 20 (Curran Associates, Inc., 2007).

42. Tancik, M. *et al.* Fourier Features Let Networks Learn High Frequency Functions in Low Dimensional Domains. in *Advances in Neural Information Processing Systems* (eds Larochelle, H., Ranzato, M., Hadsell, R., Balcan, M. F. & Lin, H.) vol. 33 7537–7547 (Curran Associates, Inc., 2020).

43. Kim, N. *et al.* Single-cell RNA sequencing demonstrates the molecular and cellular reprogramming of metastatic lung adenocarcinoma. *Nat. Commun.* **11**, 2285 (2020).

44. Srivatsan, S. R. *et al.* Embryo-scale, single-cell spatial transcriptomics. *Science* **373**, 111–117 (2021).

# Supplementary information

Resolving cell states in spatial transcriptomics with SCATTER.

The primary analyses of the six pseudo-spot benchmarks used random seed 0. For each benchmark, MERFISH, Xenium or Stereo-seq cells were aggregated into spatial bins to generate mixed counts. Cell-type-specific expression targets were calculated from the same source cells; for Stereo-seq, the targets used the half of each cell's transcripts withheld from the mixed counts. Datasets and sizes: whole-brain MERFISH coronal section (80 cell types, 500 genes), hypothalamic preoptic MERFISH (9 cell types, 134 genes), human breast Xenium (19 cell types, 313 genes), and human multiple-sclerosis brain MERFISH under a cross-modality single-nucleus reference (7 cell types, 477 genes) and a held-out same-modality reference (7 cell types, 500 genes), and adult mouse hemibrain Stereo-seq (7 cell types, 2,935 genes).

# Supplementary figures

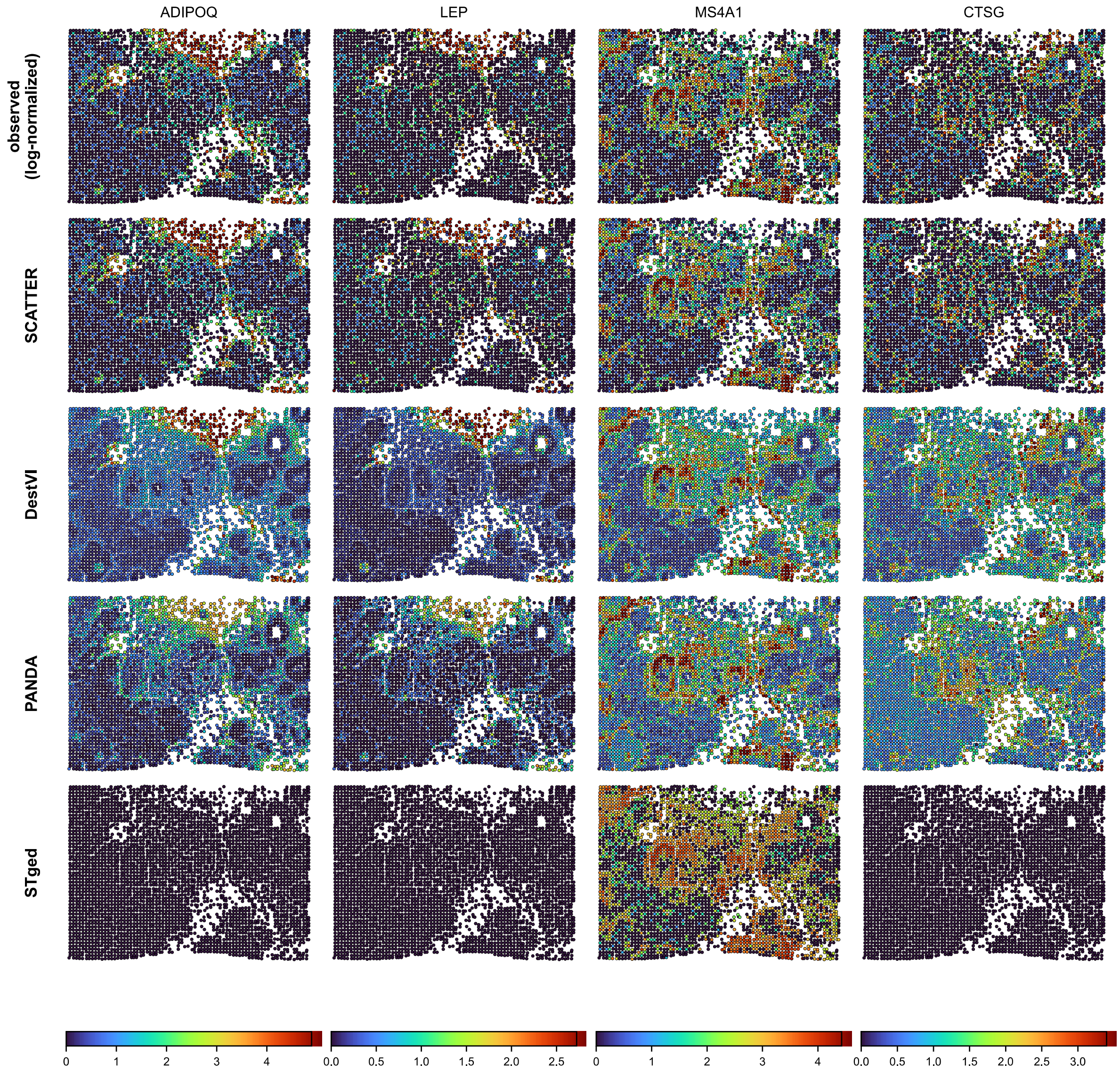

**Supplementary Figure 1. Mixed-spot expression reconstruction in breast-cancer Xenium pseudo-spots.** Columns show ADIPOQ, LEP, MS4A1 and CTSG. The top row shows observed counts normalized by each spot's library size; subsequent rows show expression reconstructed as the sum of inferred cell-type proportions multiplied by the corresponding inferred gene rates for SCATTER, DestVI, PANDA and STged. Observed and reconstructed rates are displayed as log1p(10,000 × rate). Each gene uses a common scale across methods, capped at the pooled 99th percentile.

Supplementary Fig. 1 illustrates the observed and reconstructed mixed-spot expression maps for ADIPOQ, LEP, MS4A1 and CTSG in the breast cancer Xenium benchmark. This comparison evaluates agreement between the observed spot signal and the signal reconstructed from each method's inferred cell-type proportions and cell-type-specific expression. Within-type expression recovery was evaluated separately against the cell-resolved benchmark targets.

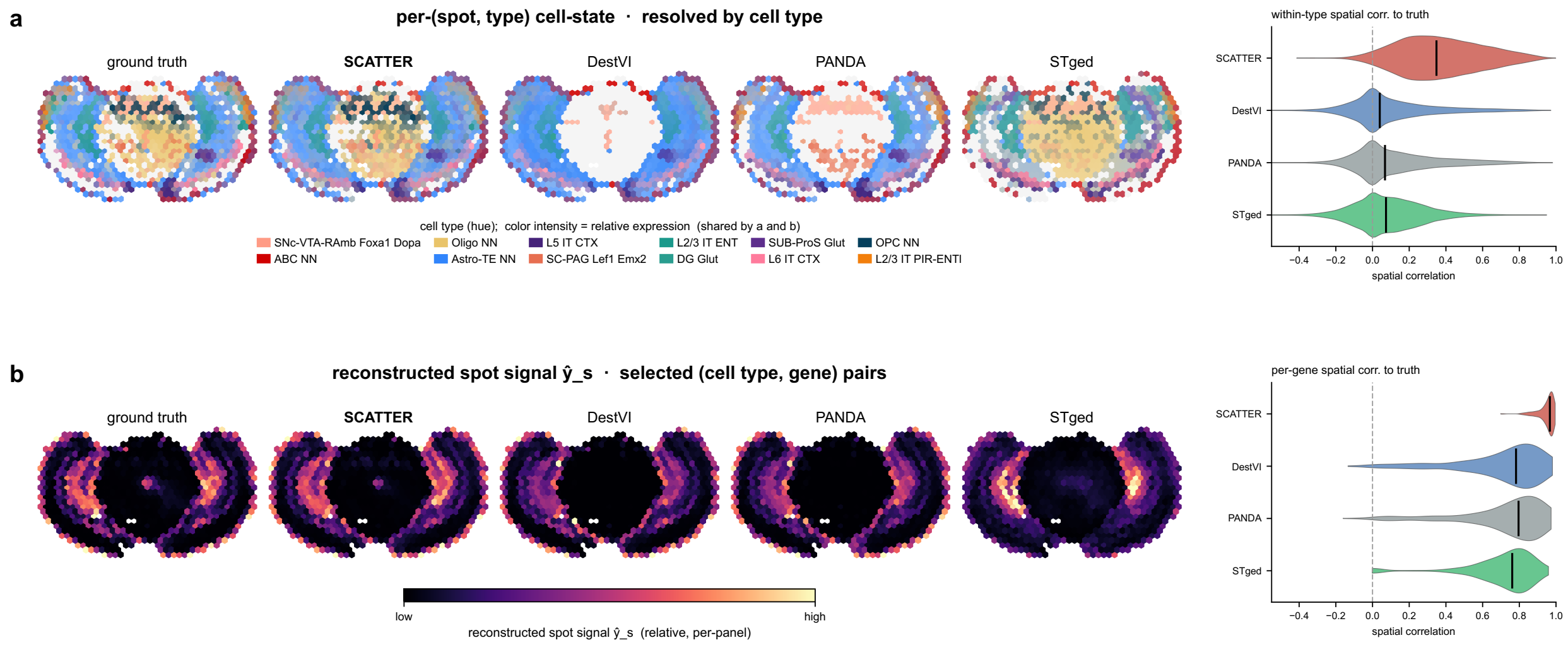

**Supplementary Figure 2. Cell-type-specific expression and reconstructed spot signals in whole-brain MERFISH pseudo-spots.** a, Twelve illustrative gene–cell-type fields, shown as hue-blended spatial maps for source-cell targets and SCATTER, DestVI, PANDA and STged. Each hexagon blends the hues of cell types present at that location, with intensity representing relative within-type expression. Examples were selected for target coverage, high SCATTER recovery correlations and spatial spread. The violin plot summarizes within-type spatial Pearson correlations across eligible gene–cell-type pairs. b, Spot signals reconstructed from selected gene–cell-type pairs using true cell-type proportions, with intensity scaled within each map. The violin plot summarizes per-gene spatial correlations across all genes between reconstructed and benchmark-target mixed-spot expression, both calculated using true cell-type proportions. For a, violin eligibility requires true cell-type proportion >0.05 in at least 50 spots and nonzero target expression in >15% of those spots. In both violins, undefined correlation coefficients are displayed as zero.

The whole-brain MERFISH comparison showed that SCATTER more closely reproduced the benchmark-target cell-type-specific expression maps across the twelve illustrated cell types (Supplementary Fig. 2a). SCATTER also achieved higher within-type spatial correlations and higher per-gene spatial correlations of reconstructed spot expression than DestVI, PANDA and STged (Supplementary Fig. 2a,b).

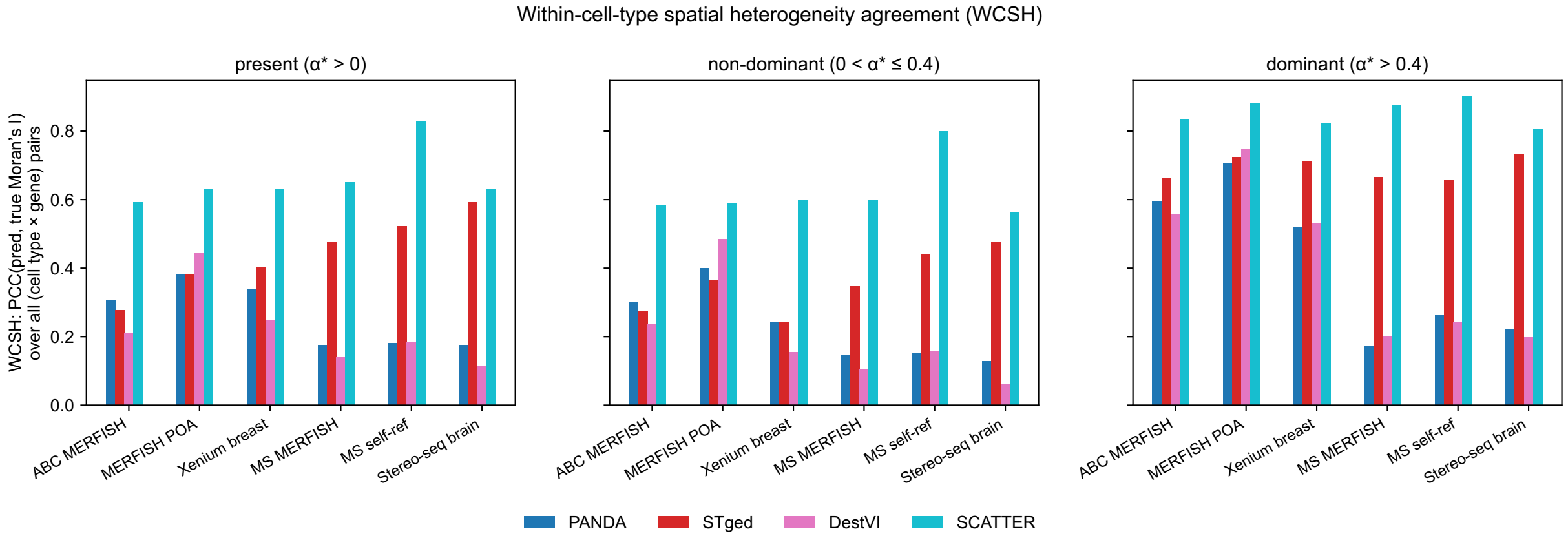

**Supplementary Figure 3. Agreement of within-type spatial autocorrelation with benchmark targets.** Pearson correlations between predicted and target Moran's I across gene–cell-type pairs in six benchmarks. Panels distinguish present ($\alpha^* > 0$), non-dominant ($0 < \alpha^* \leq 0.4$) and dominant ($\alpha^* > 0.4$) subsets. SCATTER showed the highest correlations across datasets and subsets. $\alpha^*$ denotes the true cell-type proportion at a spot. ABC MERFISH and MERFISH POA denote the Allen Brain Cell Atlas whole-brain and hypothalamic preoptic-area datasets, respectively. MS MERFISH and MS self-ref use the same multiple sclerosis MERFISH cohort, with a single-nucleus RNA-seq reference and a reference drawn from MERFISH cells excluded from pseudo-spot construction, respectively.

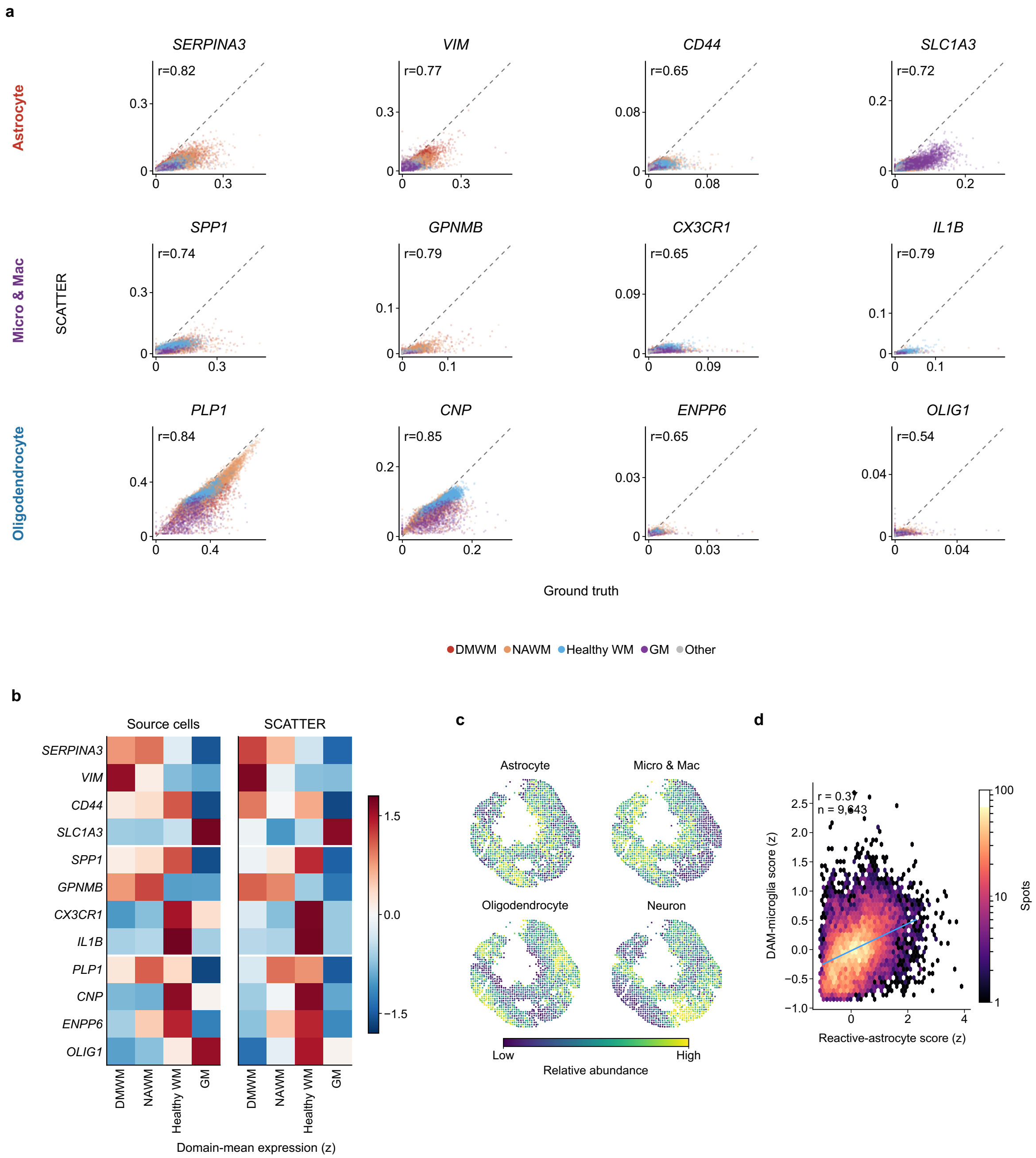

**Supplementary Figure 4. Glial expression recovery and tissue-domain context in the multiple-sclerosis MERFISH benchmark.** a, Recovered versus source-cell expression for twelve illustrative gene–cell-type pairs. Dots are eligible pseudo-spots colored by tissue domain; dashed lines indicate y = x and r denotes cohort-wide Pearson correlation. b, Source-derived and recovered domain-mean expression, separately z-scored across four structural domains for each gene on a shared color scale. c, Source-derived cell-type proportions in the representative section, percentile-ranked within each map. d, Recovered reactive-astrocyte versus disease-associated-microglia (DAM) scores across 9,643 cohort pseudo-spots containing >5% of both cell types. Hexagon color indicates spot counts; the line is a linear fit. Program definitions and scoring are described in Supplementary Method 3. DMWM denotes demyelinated white matter, NAWM denotes normal-appearing white matter, WM denotes white matter and GM denotes gray matter. “Micro & Mac” denotes microglia and macrophages.

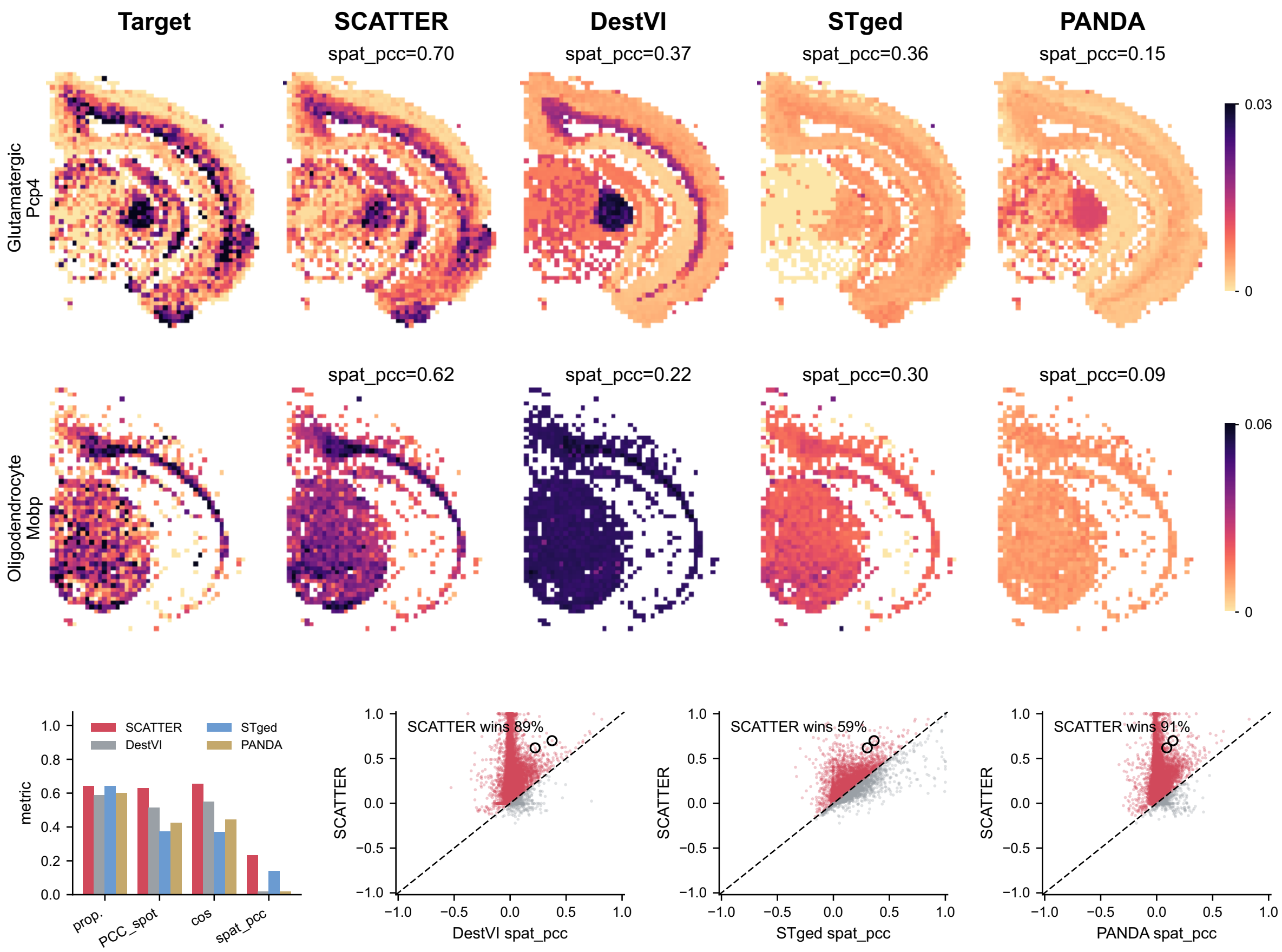

**Supplementary Figure 5. Recovery of spatial expression in mouse hemibrain Stereo-seq pseudo-spots.** Targets were constructed from transcripts withheld from the mixed counts. Maps show Pcp4 in glutamatergic neurons and Mobp in oligodendrocytes for the target and all four methods, using a common color scale within each row. Map labels give within-cell-type spatial Pearson correlations. Bars report the four summary metrics defined in Fig. 2. Scatter plots compare SCATTER with DestVI, STged and PANDA across gene–cell type pairs with defined correlations for both methods and at least 30 pseudo-spots where the true cell-type proportion exceeded 0.05. Red points favor SCATTER, open circles mark the mapped examples and dashed diagonals indicate equality. The corresponding win rates are 89.4%, 58.8% and 91.1%, respectively.

# Supplementary tables

**Supplementary Table 1. Accuracy of cell-type proportions.**

Whole-brain MERFISH

| **method** | **pcc_ct** | **spear_ct** | **cos_spot** | **jsd_spot** | **rmse** | **nmi** |
|---|---|---|---|---|---|---|
| NNLS | 0.613 | 0.144 | 0.492 | 0.332 | 0.058 | 0.496 |
| STINR | 0.448 | 0.138 | 0.705 | 0.224 | 0.041 | 0.617 |
| PANDA | 0.865 | 0.230 | 0.875 | 0.122 | 0.028 | 0.713 |
| STged | 0.809 | 0.072 | 0.920 | 0.089 | 0.025 | 0.727 |
| DestVI | 0.628 | 0.188 | 0.786 | 0.178 | 0.035 | 0.697 |
| Cell2location | 0.872 | **0.256** | 0.931 | **0.064** | **0.021** | 0.727 |
| **SCATTER** | **0.890** | 0.233 | **0.939** | 0.082 | **0.021** | **0.782** |

Hypothalamic MERFISH

| method | pcc_ct | spear_ct | cos_spot | jsd_spot | rmse | nmi |
|---|---|---|---|---|---|---|
| NNLS | 0.785 | 0.587 | 0.888 | 0.092 | 0.098 | 0.525 |
| STINR | 0.759 | 0.521 | 0.845 | 0.091 | 0.113 | 0.467 |
| PANDA | 0.774 | 0.579 | 0.917 | 0.075 | 0.085 | 0.527 |
| STged | 0.831 | 0.527 | 0.936 | 0.062 | 0.078 | 0.616 |
| DestVI | 0.810 | 0.598 | 0.910 | 0.096 | 0.089 | 0.601 |
| Cell2location | **0.872** | **0.662** | 0.941 | **0.047** | 0.071 | 0.616 |
| **SCATTER** | 0.866 | 0.658 | **0.950** | 0.052 | **0.067** | **0.617** |

Breast cancer Xenium

| method | pcc_ct | spear_ct | cos_spot | jsd_spot | rmse | nmi |
|---|---|---|---|---|---|---|
| NNLS | 0.534 | 0.346 | 0.668 | 0.231 | 0.116 | 0.493 |
| STINR | 0.350 | 0.198 | 0.618 | 0.266 | 0.142 | 0.455 |
| PANDA | 0.670 | 0.377 | 0.869 | 0.132 | 0.085 | 0.588 |
| STged | 0.689 | 0.314 | 0.748 | 0.176 | 0.111 | 0.496 |
| DestVI | **0.736** | 0.381 | **0.878** | **0.128** | 0.083 | **0.644** |
| Cell2location | 0.688 | **0.412** | 0.755 | 0.162 | 0.106 | 0.496 |
| **SCATTER** | 0.730 | 0.397 | 0.869 | 0.131 | **0.082** | 0.619 |

MS brain (cross-modality)

| method | pcc_ct | spear_ct | cos_spot | jsd_spot | rmse | nmi |
|---|---|---|---|---|---|---|
| NNLS | 0.567 | 0.427 | 0.478 | 0.317 | 0.268 | 0.052 |
| STINR | 0.357 | 0.288 | 0.723 | 0.195 | 0.183 | 0.217 |
| PANDA | 0.696 | 0.588 | 0.906 | 0.091 | 0.109 | 0.524 |
| STged | 0.837 | 0.568 | 0.948 | 0.063 | 0.102 | 0.629 |
| DestVI | 0.825 | 0.713 | 0.955 | 0.036 | 0.080 | 0.607 |
| Cell2location | 0.853 | **0.730** | 0.952 | 0.044 | 0.092 | 0.629 |
| **SCATTER** | **0.858** | 0.721 | **0.968** | **0.035** | **0.069** | **0.659** |

MS brain (same-modality)

| method | pcc_ct | spear_ct | cos_spot | jsd_spot | rmse | nmi |
|---|---|---|---|---|---|---|
| NNLS | 0.758 | 0.553 | 0.939 | 0.065 | 0.101 | 0.593 |
| STINR | 0.534 | 0.417 | 0.891 | 0.089 | 0.127 | 0.430 |
| PANDA | 0.803 | 0.705 | 0.959 | 0.038 | 0.076 | 0.641 |
| STged | 0.872 | 0.613 | 0.966 | 0.042 | 0.076 | 0.653 |
| DestVI | 0.865 | 0.704 | 0.971 | 0.029 | **0.065** | 0.659 |
| Cell2location | 0.883 | **0.725** | 0.968 | 0.030 | 0.070 | 0.653 |
| **SCATTER** | **0.887** | 0.715 | **0.972** | **0.027** | **0.065** | **0.667** |

Mouse hemibrain (Stereo-seq)

| method | pcc_ct | spear_ct | cos_spot | jsd_spot | rmse | nmi |
|---|---|---|---|---|---|---|
| NNLS | 0.478 | 0.352 | 0.822 | 0.153 | 0.174 | 0.424 |
| STINR | 0.498 | 0.358 | 0.846 | 0.102 | 0.138 | 0.372 |
| PANDA | 0.600 | 0.478 | 0.909 | 0.073 | 0.107 | 0.496 |
| STged | 0.641 | 0.489 | 0.907 | 0.079 | 0.108 | 0.505 |
| DestVI | 0.588 | 0.467 | 0.888 | 0.095 | 0.117 | 0.409 |
| cell2location | **0.646** | **0.541** | 0.910 | 0.075 | 0.105 | 0.505 |
| **SCATTER** | 0.643 | 0.534 | **0.917** | **0.072** | **0.100** | **0.509** |

Per-cell-type Pearson (pcc_ct) and Spearman (spear_ct) correlation of predicted against true proportions, per-spot cosine similarity between proportion vectors (cos_spot), per-spot Jensen-Shannon divergence (jsd_spot, lower is better), proportion root-mean-square error (rmse, lower is better), and agreement of each spot's dominant (argmax) cell-type label between prediction and truth by normalized mutual information (nmi, higher is better). Best per column in bold. Cell2location supplies SCATTER's proportion prior and appears here as a baseline.

SCATTER achieved the highest per-cell-type Pearson correlation for cell-type proportions on whole-brain MERFISH and both multiple-sclerosis benchmarks. On hypothalamic MERFISH, its correlation was close to that of its Cell2location prior. On breast Xenium, which used a disjoint cross-assay reference, SCATTER's correlation for cell-type proportions was close to that of DestVI, the strongest baseline for this metric. On Stereo-seq mouse brain, its correlation was close to that of Cell2location (0.643 versus 0.646) and its RMSE was the lowest.

**Supplementary Table 2. Per-(spot, cell-type) expression accuracy.**

| dataset | method | PCC_spot | cos_sk | jsd_sk | spatPCC |
|---|---|---|---|---|---|
| Whole-brain MERFISH | PANDA | 0.608 | 0.627 | 0.218 | 0.143 |
| | STged | 0.457 | 0.352 | 0.146 | 0.094 |
| | DestVI | 0.808 | 0.825 | 0.144 | 0.081 |
| | **SCATTER** | **0.848** | **0.860** | **0.117** | **0.356** |
| Hypothalamic MERFISH | PANDA | 0.630 | 0.675 | 0.210 | 0.083 |
| | STged | 0.520 | 0.543 | 0.180 | 0.045 |
| | DestVI | 0.797 | 0.823 | 0.141 | 0.116 |
| | **SCATTER** | **0.863** | **0.877** | **0.114** | **0.290** |
| Breast cancer Xenium | PANDA | 0.664 | 0.684 | 0.194 | 0.062 |
| | STged | 0.597 | 0.446 | **0.139** | 0.094 |
| | DestVI | 0.724 | 0.761 | 0.160 | 0.091 |
| | **SCATTER** | **0.779** | **0.795** | 0.147 | **0.287** |
| MS brain (cross-modality) | PANDA | 0.470 | 0.402 | 0.368 | 0.034 |
| | STged | 0.515 | 0.386 | 0.207 | 0.147 |
| | DestVI | **0.782** | **0.756** | 0.200 | 0.058 |
| | **SCATTER** | 0.745 | 0.722 | **0.196** | **0.276** |
| MS brain (same-modality) | PANDA | 0.848 | 0.826 | 0.171 | 0.074 |
| | STged | 0.621 | 0.524 | 0.185 | 0.149 |
| | DestVI | **0.867** | **0.840** | **0.163** | 0.099 |

| dataset | method | PCC_spot | cos_sk | jsd_sk | spatPCC |
|---|---|---|---|---|---|
| | **SCATTER** | 0.830 | 0.799 | 0.173 | **0.274** |
| Mouse hemibrain (Stereo-seq) | PANDA | 0.425 | 0.444 | 0.389 | 0.019 |
| | STged | 0.374 | 0.369 | 0.329 | 0.139 |
| | DestVI | 0.513 | 0.550 | 0.354 | 0.017 |
| | **SCATTER** | **0.630** | **0.655** | **0.300** | **0.233** |

Per-spot expression Pearson (PCC_spot, the within-spot correlation between predicted and true expression over the present cell-type and gene entries, distinct from the spot-level reconstruction correlation), per-(spot, cell-type) cosine (cos_sk), per-(spot, cell-type) Jensen-Shannon divergence (jsd_sk, lower is better), and the within-type spatial correlation to ground truth (spatPCC, true proportion >0.05 in at least three spots and finite correlation, without the gene-frequency filter used for Supplementary Table 3). Static-signature methods (NNLS, STINR) return a constant profile per cell type and are omitted here; the four expression-recovery methods are shown, best per column in bold.

SCATTER achieved the highest mean within-type spatial correlation in all six benchmarks. SCATTER had the highest per-spot expression correlation and per-(spot, cell-type) cosine similarity in four settings. DestVI led on both metrics in the two multiple sclerosis settings. SCATTER had the lowest per-(spot, cell-type) Jensen–Shannon divergence in four settings. STged had the lowest divergence in breast cancer Xenium, and DestVI had the lowest divergence in the same-modality multiple sclerosis benchmark (Supplementary Table 2).

**Supplementary Table 3. Paired comparisons of within-type spatial-expression recovery.**

| dataset | baseline | pairs | SCATTER win % | median spatPCC | BH q |
|---|---|---|---|---|---|
| Whole-brain MERFISH | DestVI | 29,636 | 84.4 | **0.365** vs 0.062 | $< 10^{-300}$ |
| | PANDA | 30,587 | 80.6 | **0.364** vs 0.105 | $< 10^{-300}$ |
| | STged | 31,888 | 82.4 | **0.357** vs 0.082 | $< 10^{-300}$ |
| Hypothalamic MERFISH | DestVI | 1,151 | 79.6 | **0.262** vs 0.077 | $1.9{\times}10^{-98}$ |
| | PANDA | 1,053 | 85.2 | **0.259** vs 0.048 | $8.2{\times}10^{-123}$ |
| | STged | 1,002 | 84.5 | **0.262** vs 0.041 | $1.2{\times}10^{-108}$ |
| Breast cancer Xenium | DestVI | 3,982 | 88.5 | **0.287** vs 0.085 | $< 10^{-300}$ |
| | PANDA | 3,914 | 92.8 | **0.286** vs 0.046 | $< 10^{-300}$ |
| | STged | 3,658 | 91.5 | **0.287** vs 0.077 | $< 10^{-300}$ |
| MS brain (cross-modality) | DestVI | 1,332 | 96.9 | **0.325** vs 0.074 | $1.2{\times}10^{-214}$ |
| | PANDA | 1,331 | 98.9 | **0.324** vs 0.036 | $8.9{\times}10^{-219}$ |
| | STged | 1,287 | 97.0 | **0.318** vs 0.158 | $3.7{\times}10^{-208}$ |
| MS brain (same-modality) | DestVI | 1,240 | 91.0 | **0.329** vs 0.153 | $1.1{\times}10^{-180}$ |
| | PANDA | 1,240 | 96.6 | **0.329** vs 0.116 | $1.3{\times}10^{-198}$ |
| | STged | 1,230 | 97.1 | **0.327** vs 0.174 | $1.3{\times}10^{-199}$ |
| Mouse hemibrain (Stereo-seq) | DestVI | 1,526 | 96.3 | **0.234** vs 0.054 | $9.0{\times}10^{-238}$ |
| | PANDA | 1,522 | 96.1 | **0.233** vs 0.057 | $2.0{\times}10^{-236}$ |
| | STged | 1,512 | 81.0 | **0.233** vs 0.143 | $2.5{\times}10^{-160}$ |

SCATTER was compared with each expression-recovery baseline using one-sided paired Wilcoxon signed-rank tests on gene–cell-type spatial Pearson correlations. Eligible pairs required at least 20 spots with true cell-type proportion >0.05 and nonzero target expression in >15% of those spots. Benjamini–Hochberg adjustment was applied across all 18 dataset–baseline comparisons. Win rate is the percentage of evaluated pairs with a higher within-type spatial correlation for SCATTER than for the baseline; median values are SCATTER versus baseline. For breast Xenium, SCATTER exceeded DestVI in within-type spatial Spearman correlation for 88.5% of 4,025 evaluated pairs.

**Supplementary Table 4. Within-type spatial-autocorrelation agreement.**

| dataset | PANDA | STged | DestVI | SCATTER |
|---|---|---|---|---|
| Whole-brain MERFISH | 0.401 | 0.405 | 0.335 | **0.672** |
| Hypothalamic MERFISH | 0.496 | 0.490 | 0.559 | **0.700** |
| Breast cancer Xenium | 0.367 | 0.452 | 0.311 | **0.685** |
| MS brain (cross-modality) | 0.165 | 0.497 | 0.148 | **0.710** |
| MS brain (same-modality) | 0.199 | 0.540 | 0.194 | **0.843** |
| Mouse hemibrain (Stereo-seq) | 0.175 | 0.602 | 0.125 | **0.667** |

Unweighted mean of the Pearson correlations between predicted and true within-type spatial autocorrelation (Moran's I) in three cell-type proportion strata: present, non-dominant and dominant. Each subset uses an eight-nearest-neighbor, row-normalized spatial weight matrix on its selected spots, and the correlation is taken over gene–cell-type pairs with finite values. PANDA, STged and DestVI are included because they return location-specific cell-type expression profiles. NNLS and STINR emit a constant profile per type and have no within-type variation to measure, whereas Cell2location does not return a spot-by-cell-type expression tensor.

SCATTER achieved the highest agreement on every dataset (0.67–0.84), compared with 0.41–0.60 for the strongest of PANDA, STged and DestVI on each dataset. The largest margin over the strongest baseline was on the same-modality MS benchmark (0.30).

**Supplementary Table 5. Marker concordance on native human tumor tissue.**

| dataset | method | marker concordance | K_scored/K_total |
|---|---|---|---|
| Melanoma | **SCATTER** | **0.639** | 7/7 |
| | Cell2location | 0.634 | 7/7 |
| | PANDA | 0.579 | 7/7 |
| | STged | 0.568 | 7/7 |
| | DestVI | 0.230 | 7/7 |
| | NNLS | 0.186 | 5/7 |
| Squamous-cell carcinoma | PANDA | 0.466 | 6/11 |
| | Cell2location | 0.467 | 6/11 |
| | SCATTER | 0.461 | 6/11 |
| | **DestVI** | **0.500** | 6/11 |
| | STged | 0.476 | 6/11 |
| | NNLS | 0.393 | 5/11 |

Values are mean across-type Spearman correlations between inferred cell-type proportions and observed marker-program scores across spots. K_scored/K_total gives the number of scored cell types relative to all reference cell types; eligibility criteria are defined in Supplementary Note 2. Bold indicates the highest concordance per dataset.

SCATTER achieved the highest marker concordance on melanoma (0.639), slightly above Cell2location (0.634). On squamous-cell carcinoma, DestVI had the highest marker concordance (0.500), followed by STged (0.476), Cell2location (0.467), PANDA (0.466), SCATTER (0.461) and NNLS (0.393).

**Supplementary Table 6. Baseline settings and deviations from published workflows for the benchmark comparisons.**

| method | key settings | deviation from published workflow |
|---|---|---|
| Cell2location, DestVI | Cell2location: N_cells_per_location = 8, detection_alpha = 20; 250 reference-training epochs and 30,000 spatial-training epochs. DestVI: 300 single-cell and 2,500 spatial-training epochs. | none |
| NNLS | reference-signature non-negative least squares | none (static-signature control) |
| STINR | eight-nearest-neighbor spatial graph; 2,000 training steps | spatial coordinates rescaled to STINR's coordinate calibration (its default assumes DLPFC pixel coordinates); training steps reduced from the upstream default of 14,001 to 2,000 |

| method | key settings | deviation from published workflow |
|---|---|---|
| PANDA | replication_mode = fair | PANDA's paper-mode preprocessing (per-cell filter, cell-type-size downsampling, 5,000-gene spatial-HVG selection) is disabled so it scores on the shared gene panel |
| STged | reference total-count floor nUMI = 1; maxiter = 100; spatial lattice Hex (Square for the datasets of Supplementary Note 4) | STged's default floor nUMI = 100 (tuned for whole-transcriptome scRNA) is lowered to 1 so every method scores against the same reference set |

The baseline methods used random seed 0 and the shared gene panel. Cell2location supplied SCATTER's training prior. STged received Cell2location proportions for the pseudo-spot benchmarks and melanoma, and the source study's deposited EnDecon proportions for squamous-cell carcinoma. The wrapper restricted these inputs to the cell types retained by STged and renormalized them before expression fitting. For composition scoring, it mapped the returned proportion matrix back to the original spot and cell-type axes and renormalized the mapped proportions.

**Supplementary Table 7. Dataset dimensions.**

| dataset | role | spots | cell types | genes | reference cells | aggregated cells |
|---|---|---|---|---|---|---|
| Whole-brain MERFISH | pseudo-spot benchmark | 2,054 | 80 | 500 | 50,000 | 88,784 |
| Hypothalamic MERFISH | pseudo-spot benchmark | 321 | 9 | 134 | 29,756 | 5,343 |
| Breast cancer Xenium | pseudo-spot benchmark | 3,724 | 19 | 313 | 22,271 | 158,179 |
| MS brain (cross-modality) | pseudo-spot benchmark | 18,828 | 7 | 477 | 45,242 | 367,004 |
| MS brain (same-modality) | pseudo-spot benchmark | 15,783 | 7 | 500 | 44,618 | 251,093 |
| Mouse hemibrain (Stereo-seq) | pseudo-spot benchmark | 2,209 | 7 | 2,935 | 9,125 | 43,075 |
| Melanoma | native comparison | 292 | 7 | 14,520 | 4,134 | n/a |
| Squamous-cell carcinoma | native comparison | 646 | 11 | 12,005 | 2,683 | n/a |
| Colorectal cancer | native application | 41,400 | 18 | 2,032 | 8,963 | n/a |
| Breast Visium CytAssist | native application | 4,963 | 19 | 5,012 | 22,271 | n/a |

Processed dimensions of each dataset: spatial spots, cell types, genes on the shared panel, single-cell reference cells, and, for the pseudo-spot benchmarks with a cell-resolved ground truth, the total single cells aggregated into the spots. The native sequencing-based datasets have no cell-resolved expression targets (aggregated cell counts, not applicable).

**Supplementary Table 8. Default SCATTER architecture, training and backfitting settings.**

| Parameter | Default setting |
|---|---|
| **Architecture** | |
| Spot and reference hidden dimension | $d_h = 256$ |
| Coupling embedding dimension | $d_g = 128$ |
| Latent-code dimension | $d_z = 8$ |
| Graph-transformer depth | 1 |
| Attention heads | 4 |

| Parameter | Default setting |
| --- | --- |
| Fourier frequencies | 32 |
| Fourier scale | 5 |
| Spatial neighbors, $k$ | 8 |
| Coupling temperature, $\tau$ | Learned independently in each stage; initialized at 1.0; softplus parameterization |
| **Training** | |
| Optimizer | Adam; independent optimizer in each stage |
| Training steps | 3,000 per stage (6,000 total) |
| Reference coupling | Stage 1: softmax; stage 2: Taylor2 |
| Stage-2 composition | Fixed to the final stage-1 estimates |
| Stage-2 initialization | Fresh initialization using the same seed as stage 1 |
| Initial learning rate | 0.002 |
| Reduced learning rate | 0.0005 |
| Learning-rate switch step | 2,400 within each stage |
| Gradient-norm clipping threshold | 5 |
| Reconstruction loss | Squared error on log1p counts |
| Reconstruction-loss weight | $\lambda_{rec} = 5$ |
| Encoder-stability weight | $\lambda_{enc} = 10^{-2}$ |
| Coordinate-jitter standard deviation | $\sigma = 0.02$ |
| Latent-smoothness weight | $\lambda_z = 10^{-2}$ |
| Initial composition-prior weight | 1.0 (stage 1) |
| Composition-prior weight floor | 0.1 (stage 1) |
| Composition-prior annealing horizon | 3,000 steps; linear decay during stage 1 |
| **Backfitting** | |
| Update blend | 0.15 |
| Relative-change tolerance | 0.01 |
| Maximum sweeps | 16 |

**Supplementary Table 9. Detailed comparisons for lung Visium HD sections evaluated against Xenium measurements.**

| Visium HD workflow | Section pairing | Xenium panel | Genes scored | proportion RMSE | within-type spatial PCC | SCATTER win rate (%) |
|---|---|---|---|---|---|---|
| After Xenium | Same section | v1 | 175 | **0.085** / 0.097 | **0.170** / 0.095 / 0.063 / 0.047 | 70.8 / 71.9 / 75.4 |
| Visium HD only | Adjacent section | v1 | 175 | **0.092** / 0.101 | **0.171** / 0.094 / 0.063 / 0.044 | 72.0 / 71.1 / 75.4 |
| Visium HD only | Adjacent section | Prime 5K | 4,441 | **0.107** / 0.116 | **0.055** / 0.029 / 0.024 / 0.011 | 60.6 / 58.4 / 62.7 |

The section labeled "After Xenium" was assayed with Xenium v1 before Visium HD. For the two "Visium HD only" sections, the indicated Xenium panels were applied to adjacent sections from the same tissue block. Proportion RMSE over 128-μm squares with labelled Xenium cells, SCATTER versus the best other method (PANDA for the post-Xenium section, Cell2location otherwise). Within-type spatial PCC lists SCATTER / STged / PANDA / DestVI, using whole-cell Xenium profiles, squares whose Xenium cell-type fraction exceeded 0.05, gene–cell type pairs computable for all methods and annotation marker genes excluded. SCATTER win % lists the comparisons with STged / PANDA / DestVI over pairs with at least 20 eligible squares and nonzero target expression in >15% of them; all two-sided paired Wilcoxon tests had $q < 10^{-33}$.

**Supplementary Table 10. Cell-type proportion and expression estimates in sci-Space embryonic cortex pseudo-spots.**

| Mixture | Reference source | proportion PCC | within-type spatial PCC | SCATTER win rate (%) | Neuron PCC (DestVI metric) |
|---|---|---|---|---|---|
| Original | Held-out slides | 0.766 / **0.807** | **0.216** / 0.172 / 0.072 / 0.046 | 70.8 / 81.8 / 86.7 | **0.50** / 0.15 |
| Original | Pseudo-spot cells | 0.772 / **0.835** | **0.214** / 0.179 / 0.115 / 0.088 | 68.4 / 71.9 / 76.6 | **0.50** / 0.38 |
| Non-neuronal upweighting | Held-out slides | 0.847 / **0.855** | **0.353** / 0.260 / 0.089 / 0.057 | 82.7 / 87.9 / 89.1 | **0.59** / 0.19 |
| Non-neuronal upweighting | Pseudo-spot cells | 0.881 / **0.885** | **0.371** / 0.266 / 0.142 / 0.105 | 85.0 / 84.7 / 84.0 | **0.60** / 0.46 |

The original mixtures use unweighted cortex cells. For non-neuronal upweighting, each non-neuronal cell contributes fivefold when constructing the pseudo-spots, and all-neuron pseudo-spots are excluded. "Pseudo-spot cells" denotes the reference drawn from the constituent cells, and "held-out slides" denotes the reference drawn from cortex cells on separate slides. Proportion PCC is SCATTER versus PANDA, the best other method in every setting. Within-type spatial PCC (SCATTER / STged / PANDA / DestVI) and SCATTER win % (versus STged / PANDA / DestVI) follow Supplementary Table 9, with true proportions; all two-sided paired Wilcoxon tests had $q < 10^{-167}$. The DestVI metric follows the DestVI benchmark code: neurons in spots with true proportion >0.2, 13 genes of interest, gene-averaged Pearson correlation with undefined values set to zero (SCATTER / DestVI).

# Supplementary notes

## Supplementary Note 1. Baseline methods

We compared cell-type proportions and expression recovery using different sets of methods. The proportion comparison included NNLS, STINR[1], Cell2location[2], DestVI[3], PANDA[4], STged[5] and SCATTER. NNLS estimated proportions from reference-derived signatures by non-negative least squares. STINR estimated spatially varying proportions against a fixed cell-type expression basis. Cell2location was the dedicated probabilistic deconvolution baseline and also supplied SCATTER's proportion prior. DestVI, PANDA, STged and SCATTER returned spatially varying cell-type-specific expression profiles and were included in the expression-recovery and spatial-autocorrelation comparisons. NNLS and STINR used a constant expression profile for each cell type, while Cell2location returned proportions without a spot-by-cell-type expression tensor. All methods were run on the shared gene panel.

Before expression evaluation, each predicted spot–cell-type gene profile was normalized to sum to one; profiles with zero total expression were retained as zero.

## Supplementary Note 2. Evaluation metrics

**Notation.** A dataset has $S$ spots, $K$ cell types and $G$ genes. Write $\alpha_{s,k}$ and $\alpha^{\star}_{s,k}$ for the predicted and true proportion of type $k$ at spot $s$, and $\hat{E}_{s,k,g}$ and $E^{\star}_{s,k,g}$ for the predicted and true expression of gene $g$ in type $k$ at spot $s$, each on the per-(spot, cell-type) rate scale. Type $k$ is present at spot $s$ when its true number of contributing cells is positive (or $\alpha^{\star}_{s,k} > 0$ when cell counts are unavailable); the expression metrics are scored only on present $(s, k)$ pairs, because a type absent from a spot has no meaningful ground-truth profile there. corr denotes Pearson correlation, $\text{corr}^{\text{S}}$ Spearman correlation and JSD the Jensen-Shannon divergence. Every metric is defined per element below, and each result reports its mean over the retained elements.

### Cell-type proportions

Proportion metrics compare predicted and true cell-type proportions across all spots.

The per-cell-type Pearson (pcc_ct) and Spearman (spear_ct) correlations test whether each type's estimated proportion tracks its true proportion across space, averaged over types:

$$\text{pcc_ct} = \frac{1}{K}\sum_{k} \text{corr}_s\left(\alpha_{\cdot,k}, \alpha^{\star}_{\cdot,k}\right), \qquad \text{spear_ct} = \frac{1}{K}\sum_{k} \text{corr}^{\text{S}}_s\left(\alpha_{\cdot,k}, \alpha^{\star}_{\cdot,k}\right),$$

Each correlation is calculated across spots for one cell type, and the resulting correlations are averaged over cell types.

The per-spot cosine (cos_spot) measures cosine similarity between the predicted and true cell-type proportion vectors at each spot:

$$\text{cos_spot} = \frac{1}{S}\sum_{s} \cos\left(\alpha_{s,\cdot}, \alpha^{\star}_{s,\cdot}\right).$$

The per-spot Jensen-Shannon divergence (jsd_spot, lower is better) is the distributional distance between the predicted and true proportion vectors:

$$\text{jsd_spot} = \frac{1}{S}\sum_{s} \text{JSD}\left(\alpha_{s,\cdot} \parallel \alpha^{\star}_{s,\cdot}\right).$$

The proportion root-mean-square error (rmse, lower is better) is the elementwise error over all (spot, type) proportions:

$$\text{rmse} = \sqrt{\frac{1}{SK}\sum_{s,k}\left(\alpha_{s,k} - \alpha^{\star}_{s,k}\right)^2}.$$

The dominant-label agreement (nmi) is the normalized mutual information between the per-spot dominant (argmax) cell-type labels of the prediction and the truth:

$$\text{nmi} = \text{NMI}\left(\left\{\underset{k}{\text{argmax}}\,\alpha_{s,k}\right\}_s, \left\{\underset{k}{\text{argmax}}\,\alpha^{\star}_{s,k}\right\}_s\right).$$

### Expression

Expression metrics compare predicted and benchmark-target profiles at spots where the evaluated cell type is present.

The per-spot expression Pearson (PCC_spot) concatenates, within each spot, the gene entries of the present types and correlates them with the truth:

$$\text{PCC_spot} = \frac{1}{S}\sum_{s} \text{corr}_{(k,g):\,k\in\mathcal{K}_s}\left(\hat{E}_{s,k,g},\, E^{\star}_{s,k,g}\right),$$

with $\mathcal{K}_s$ the set of types present at spot $s$.

The per-(spot, cell-type) cosine (cos_sk) compares predicted and true gene-expression profiles, averaged over present spot–cell-type pairs:

$$\text{cos_sk} = \underset{(s,k)\text{ present}}{\text{mean}} \cos\left(\hat{E}_{s,k,\cdot},\, E^{\star}_{s,k,\cdot}\right).$$

The per-(spot, cell-type) Jensen-Shannon divergence (jsd_sk, lower is better) is the distributional distance of the two per-type gene profiles, each normalized over genes:

$$\text{jsd_sk} = \underset{(s,k)\text{ present}}{\text{mean}} \text{JSD}\left(\hat{E}_{s,k,\cdot} \parallel E^{\star}_{s,k,\cdot}\right).$$

The within-type spatial correlation (spatPCC) measures recovery of gene-expression variation across spots within each cell type. For each cell type $k$ and gene $g$ it is the Pearson correlation, over the spots where $k$ has a true proportion above 0.05, between the recovered and true expression; the reported value averages over the scored pairs:

$$r_{k,g} = \text{corr}_s\left(\hat{E}_{s,k,g},\, E^{\star}_{s,k,g}\right), \qquad \text{spatPCC} = \underset{(k,g)}{\text{mean}}\, r_{k,g}.$$

For the mean spatPCC, a pair is scored when $\alpha^{\star}_{s,k} > 0.05$ in at least three spots and its spatial correlation is finite; no gene-expression-frequency filter is applied to this mean. Means use the finite correlations available for each method. Paired comparisons use the eligible gene–cell-type pairs shared by SCATTER and the baseline method. For paired significance tests, at least twenty such spots are required and $E^{\star}_{\cdot,k,g}$ is non-zero at more than fifteen per cent of them. A method that emits a spatially constant profile per type has no across-spot variance and an undefined $r_{k,g}$, reported as “n/a”. The main Fig. 2 scatterplots require at least 30 eligible spots per cell type without the expression-frequency filter. The comparisons with DestVI, STged and PANDA include 5,227, 4,384 and 4,994 shared gene–cell-type pairs with computable Pearson correlations, respectively. SCATTER has higher correlations for 4,643, 3,859 and 4,607 of these pairs, corresponding to win rates of 88.8%, 88.0% and 92.3% (rounded to 89%, 88% and 92% in Fig. 2). In the Stereo-seq benchmark (Supplementary Fig. 5), SCATTER has the higher correlation for 3,779 of 6,431 pairs shared with STged (58.8%). The breast-cancer comparisons in Table 1 and Supplementary Table 3 instead require at least 20 eligible spots and nonzero target expression in more than 15% of those spots; the corresponding denominators are 3,982, 3,658 and 3,914 pairs, with win rates of 88.5%, 91.5% and 92.8%. Pearson correlations are computed when the product of the centered prediction and target vector norms exceeds $10^{-12}$ for Fig. 2 or $10^{-9}$ for the paired significance tests; Supplementary Fig. 2a requires at least 50 such spots and the 15% expression filter.

**Significance**

For each dataset and expression-recovery baseline, the paired values $\{r_{k,g}\}$ of SCATTER and the baseline over their common pairs passing the twenty-spot and 15% expression filters are compared by a one-sided Wilcoxon signed-rank test; Benjamini–Hochberg correction across the family of SCATTER-against-baseline comparisons gives adjusted $q$-values. The win rate is the fraction of pairs on which SCATTER exceeds the baseline, $r_{k,g}^{\mathrm{SCATTER}} > r_{k,g}^{\mathrm{base}}$.

### Within-type spatial-autocorrelation agreement

For a within-type field $x_s = \hat{E}_{s,k,g}$ over the $N$ selected spots of type $k$, the Moran's I is

$$I(x) = \frac{N}{S_0} \frac{z^\top W z}{z^\top z}, \qquad z = x - \bar{x}, \qquad S_0 = \sum_{i,j} W_{ij},$$

with $W$ the row-standardized eight-nearest-neighbor spatial weight matrix on the selected spots. This calculation is performed separately for the present, non-dominant and dominant cell-type proportion subsets. Within each subset, agreement is the Pearson correlation, over the pairs with a finite $I$ on both, between the predicted and true within-type autocorrelations:

$$\mathrm{corr}_{(k,g)}\left(I(\hat{E}_{\cdot,k,g}),\ I(E^{\star}_{\cdot,k,g})\right),$$

**Supplementary Figure 3** reports these subset-specific correlations. Supplementary Table 4 reports the unweighted arithmetic mean of the three subset-specific correlations for each dataset. The present subset uses a positive true number of contributing cells (or true proportion >0 when cell counts are unavailable); the non-dominant subset uses 0 < true proportion ≤0.4, and the dominant subset uses true proportion >0.4.

### Marker concordance (native data)

For native sequencing-based datasets, marker concordance measures the association between inferred cell-type proportion and observed marker-gene expression. For each cell type $k$ with a curated marker set $M_k$, let $m_{s,k}$ be the mean of those markers' log-normalized observed expression in spot $s$; the concordance is the across-spot Spearman correlation between the predicted proportion of $k$ and that marker signal, averaged over the scored types:

$$\rho^{\mathrm{mark}} = \frac{1}{|\mathcal{K}'|} \sum_{k \in \mathcal{K}'} \mathrm{corr}_s^{\mathrm{S}}\left(\alpha_{s,k},\ m_{s,k}\right), \qquad m_{s,k} = \frac{1}{|M_k|} \sum_{g \in M_k} \tilde{y}_{s,g},$$

with $\mathcal{K}'$ the scored types (those with at least one marker on the panel and a non-constant proportion and marker signal). High marker concordance indicates that a cell type's inferred proportion is positively associated with its observed marker-gene expression across spots.

## Supplementary Note 3. Evaluation and downstream analyses

We evaluated cell-type proportions against the known proportions in pseudo-spot benchmarks using correlation and spot-level agreement metrics. Expression recovery was assessed by correlating inferred and source-derived expression across spots for each gene within each cell type. The primary within-type spatial-expression metric used spots where the true cell-type proportion exceeded 0.05. Paired comparisons between SCATTER and expression-recovery baselines used one-sided Wilcoxon signed-rank tests with Benjamini–Hochberg correction[6]. Agreement in within-type spatial autocorrelation was evaluated using Moran's I[7] across three cell-type proportion subsets.

For the breast-cancer Xenium benchmark, we compared inferred expression patterns and transcription-factor regulon scores with source-cell targets across tumor neighborhoods (Fig. 3). Regulon scores were calculated from measured target genes using DoRothEA regulons obtained through OmniPath[8] (Supplementary Method 1). For native breast-cancer Visium data, we defined tumor neighborhoods from inferred malignant-cell proportions and examined gene associations using three computational seeds. Partial Spearman correlations were adjusted for Cell2location-estimated proportions of the evaluated cell type and library size over the modeled gene panel (Fig. 4; Supplementary Method 2). In the multiple-sclerosis MERFISH benchmark, we compared inferred glial program scores with source-cell targets and examined their distributions across tissue domains and distances from the lesion boundary (Fig. 5; Supplementary Method 3).

For native colorectal-cancer Visium HD data, we quantified cell-type-associated expression programs around the invasion front, compared an inferred macrophage program with source-study subtype annotations, and attributed observed gene signals to cell types using proportion-weighted inferred expression. We also ranked nine curated ligand–receptor pairs using products of inferred sender and receiver expression[9] (Fig. 6; Supplementary Method 4).

## Supplementary Note 4. Additional evaluations on Visium HD lung sections and sci-Space embryo data

Three Visium HD sections from one human lung adenocarcinoma block (10x Genomics technical datasets) were evaluated against Xenium measurements of the same block (Supplementary Table 9). One Visium HD section was processed after Xenium v1 on the same slide; two Visium HD-only sections were paired with Xenium v1 or Xenium Prime 5K data from adjacent sections. We aggregated Visium HD counts into 128-μm squares and used the same 4,464 genes and eight lineages to fit all three sections. The single-cell reference included 500 cells per lineage from GEO GSE131907[10]. Xenium cells were mapped to Visium coordinates by affine registration of the hematoxylin and eosin images for the same-section pair, and by rigid registration with template matching and enhanced-correlation refinement for adjacent sections. Xenium cell-type fractions and within-type expression profiles in each square served as evaluation references. Expression scoring was restricted to genes measured by the paired Xenium panel and excluded the genes used for cell-type annotation. The Visium HD and Xenium measurements sampled different molecules and, for the adjacent-section pairs, different cells. Mean within-type spatial PCC between two disjoint halves of the Xenium cells was 0.14 for the Xenium v1 panel and 0.03 for the Xenium Prime 5K panel. These split-half comparisons measure agreement between disjoint subsets of the reference cells. Comparisons with the full Xenium reference use more cells than either half. Of the methods evaluated on these three sections, SCATTER had the lowest proportion RMSE and the highest mean within-type spatial PCC against Xenium profiles. The three whole-cell Xenium comparisons in Supplementary Table 9 used 1,072, 1,071 and 24,099 common gene–cell type pairs, respectively. Across the nine comparisons with expression-recovery baselines, the lower bounds of the 512-μm spatial-block bootstrap intervals for the mean PCC differences ranged from 0.021 to 0.120. With nuclear Xenium profiles as the reference, mean within-type spatial PCCs were 0.130, 0.125 and 0.040 for SCATTER, compared with 0.074, 0.073 and 0.024 for STged, the highest-scoring expression baseline in all three comparisons.

The sci-Space mouse embryo data[11] (GEO GSE166692) were processed following the DestVI study[3]: cortex cells of four types (neurons, radial glia, endothelial cells and connective tissue progenitors) from slides 8, 9, 11, 13 and 14, 4,000 highly variable genes, pseudo-spots formed by summing the cells at one array position and subsampling their counts with probability 0.1, and an inflated setting in which non-neuronal cells were counted five times and all-neuron spots were removed. We grouped cells by slide, row and column to preserve distinct spatial positions. In the published notebook, the product of these indices was used as the grouping key. This combined some distinct positions into 776 groups for the same 10,884 cells, compared with 1,131 occupied positions using the three indices jointly. The inflated setting retained 638 pseudo-spots. Following the published benchmark, we fitted both mixture settings using the constituent cells of the evaluated pseudo-spots as the single-cell reference. We repeated each setting using cortex cells from sections that did not contribute to the evaluated pseudo-spots. Within each of the four settings, SCATTER had the highest mean within-cell-type spatial PCC on gene–cell type pairs with defined correlations for SCATTER, DestVI, PANDA and STged. SCATTER had paired win rates of 68.4–89.1% against the three expression baselines. PANDA had the highest PCC for cell-type proportions among the seven methods evaluated in each setting (Supplementary Table 10).

# Supplementary Methods

## Supplementary Method 1. Breast-cancer Xenium microenvironment and regulon scoring

For the breast-cancer analysis (main Fig. 3), each pseudo-spot's neighborhood tumor fraction was calculated from its 15 nearest other pseudo-spots. The tumor proportion summed the source-study Invasive_Tumor, Prolif_Invasive_Tumor, DCIS_1 and DCIS_2 annotations, and the neighborhood fraction was the mean of these proportions across the neighbors. This neighborhood covariate was used to evaluate inferred and source-derived expression patterns, and was not supplied during SCATTER fitting. Gene-field and transcription-factor score comparisons were restricted to pseudo-spots in which the true proportion of the evaluated cell type exceeded 0.05. The four illustrative gene–cell-type pairs in main Fig. 3a,c were selected for biological relevance. The comparison in main Fig. 3b used pseudo-spots in the highest and lowest tertiles of neighborhood malignant cell fraction, with a two-sided Mann–Whitney U test for each gene. Differential expression was called from the inferred expression at nominal $P < 0.05$ and absolute log2 fold change above 0.4, without multiple-testing correction. Direction agreement was the fraction of these genes for which the inferred expression change matched the change calculated from source cells. The gene-by-cell-type comparison in main Fig. 3d evaluated all 75 combinations of the five displayed cell types and 15 displayed genes. The panel comprised endothelial, myoepithelial, stromal and myeloid markers present in the assay. Transcription-factor scores (main Fig. 3e–g) were calculated using DoRothEA A and B regulons[12] via OmniPath[8]. Within each cell type, a factor's score was the signed regulon-weighted mean of its assayed target genes, with each gene z-scored across eligible spots; regulons required at least five measured targets. Recovery was measured by within-type spatial Spearman correlation between inferred and source-derived scores across eligible factor–cell-type pairs.

## Supplementary Method 2. Native breast-cancer Visium preprocessing and association analysis

We analyzed native Visium CytAssist counts and registered histology from Janesick et al.[13], together with an scFFPE-seq reference sampled from tissue curls adjacent to the spatial sections in the same FFPE block. The native Visium section and the Xenium measurements used for the pseudo-spot benchmark originated from this block. After preprocessing, the analysis retained 4,963 Visium spots, 22,271 reference cells and 19 annotated cell types. The modeled panel contained 5,012 genes: the 5,000 eligible shared genes with the highest variance in reference log-normalized expression, supplemented with genes from a fixed marker list.

Visium spots were required to be labeled in tissue, contain at least 1,000 counts and 100 detected genes, and have a positive total count summed across the 307 genes shared with the preceding 313-gene Xenium panel. Reference cells required a source-study annotation and a positive total count across those 307 genes; undefined, unassigned, not-plotted and unknown labels were excluded. Reference types with at least 30 retained cells were included, with at most 2,000 cells sampled per type using a fixed preprocessing seed of 0. Genes eligible for the variance-ranked panel were observed in at least ten Visium spots and had positive total reference counts. Reference rates were scaled to 10,000 counts per cell and log1p-transformed before variance calculation.

For each random seed (0, 1 and 2), we fitted Cell2location and then fitted SCATTER using the Cell2location prior from the same seed. Downstream analyses fixed the neighborhood variable, cell-type presence masks and cell-type proportion adjustment covariates to seed 0 across all three SCATTER fits. Neighborhood tumor fraction was the mean SCATTER seed-0 malignant-cell proportion among the 15 nearest other spots, summing the invasive-tumor and proliferating invasive-tumor populations and both ductal carcinoma in situ (DCIS) populations. For each cell type, the presence mask required a Cell2location seed-0 estimated proportion greater than 0.05. This yielded 3,971 eligible stromal spots, 350 Macrophage 1 spots and 3,728 Macrophage 2 spots.

Exploratory screening included all 5,012 modeled genes across the 15 cell types represented in at least 50 eligible spots. Within each type, the upper and lower tertiles of neighborhood tumor fraction defined tumor-high and tumor-low groups. Gene expression was compared between these groups using two-sided Mann–Whitney U tests, with Benjamini–Hochberg correction applied separately within each cell type and seed. Cross-seed differential-expression candidates required a consistent direction of change, an absolute log2 fold change of at least 0.4 and $q < 0.05$ in all three seeds. Candidates were ranked by the minimum absolute log2 fold change across seeds.

We selected gene–cell-type pairs for continuous association analysis from three sources: up to ten highest-ranked upregulated and ten highest-ranked downregulated differential-expression candidates per cell type; all measured pairs from a manually curated biological gene panel; and the four illustrative gene–cell-type relationships from the preceding breast-cancer analysis. Matching the original stromal example also selected a separately annotated stromal/T-cell hybrid population, which was analyzed separately from the Stromal population shown in main Fig. 4. After deduplication, the combined set contained 442 pairs. Pairs included through the curated panel or illustrative examples did not have to satisfy the differential-expression thresholds.

Continuous associations were measured using partial Spearman correlations within each cell type's fixed presence mask. Inferred expression, neighborhood tumor fraction and the adjustment covariates were first converted to ranks. The expression ranks and neighborhood-fraction ranks were each residualized by ordinary least squares (OLS) regression on an intercept, ranked Cell2location seed-0 own-type proportion and ranked log1p library size. Library size was the sum of observed counts across the 5,012 modeled genes. The partial Spearman coefficient was the Pearson correlation between the resulting OLS residuals, which were not ranked again. Neighborhood fraction, presence masks and adjustment covariates remained fixed across seeds.

For spatial visualization, we calculated the median inferred relative expression across the three SCATTER seeds at each eligible spot and transformed it as log1p(10,000 × relative expression). For each gene, the upper color-scale limit was set to the 99th percentile of transformed median expression pooled across eligible spots from the cell types displayed for that gene. Values above this limit were assigned the uppermost color without exclusion from the underlying analysis. Individual spot values were rendered as opaque hexagonal tiles without interpolation or smoothing. Tile areas did not represent capture-disc footprints. All spatial panels used the same crop and orientation.

This analysis was exploratory, and the three random seeds represented computational repetitions on one tissue section. Differential-expression screening q-values did not account for spatial dependence. No significance tests were performed for the partial Spearman correlations.

## Supplementary Method 3. Multiple-sclerosis glial program scoring

For the multiple-sclerosis analysis (main Fig. 5), we used the cross-modality benchmark with the GSE301908 single-nucleus reference. Each glial program was scored per pseudo-spot as the mean of its marker-gene expression after gene-wise z-scoring across spots in which the corresponding cell type had a true proportion above 0.05. The programs comprised reactive astrocytes (SERPINA3, VIM, C3, CD44), homeostatic astrocytes (AQP4, SLC1A3, GJA1), disease-associated microglia (SPP1, GPNMB, TREM2, APOE, CD9, LPL, FABP5) in the microglia–macrophage compartment, and oligodendrocyte–myelin expression (PLP1, MOG, MAG, MOBP, CNP) in oligodendrocytes. Each pseudo-spot was assigned a tissue domain (demyelinated lesion, normal-appearing white matter, healthy white matter or gray matter) according to the majority BANKSY[14] label among its contributing cells. The displayed section was selected for the most balanced lesion and normal-appearing white-matter counts among astrocyte-containing spots. Within this section, signed distance to the lesion boundary was defined as the nearest-neighbor distance to the opposite domain, negative within the lesion and positive in normal-appearing white matter. Recovery across eligible cohort spots was assessed by Pearson correlation between inferred and cell-resolved source-derived program scores. We also correlated inferred reactive-astrocyte scores with the fraction of astrocytes assigned disease-associated substate labels in the source study. For the marker-by-domain comparison (Supplementary Fig. 4b), inferred and source-derived expression were separately averaged across eligible cohort spots within each of the four structural domains, and each set of four domain means was z-scored across domains for each gene. The illustrative markers in main Fig. 5c,f and Supplementary Fig. 4a,b were selected from assayed glial marker candidates based on target-expression coverage, spatial Pearson recovery and agreement between source-derived and inferred domain-mean expression patterns. All eligible pseudo-spots, including those assigned to vascular–immune regions, were retained in the expression-recovery scatter plots. Regional-mean comparisons covered the four structural domains. The marker selection did not alter the program gene sets or scores. Source-derived cell-type proportions in the representative section are shown in Supplementary Fig. 4c. The program correlation in Supplementary Fig. 4d used all cohort spots in which both astrocyte and microglia–macrophage proportions exceeded 0.05.

## Supplementary Method 4. Colorectal-cancer invasion-front and interaction scoring

For the native colorectal-cancer application (main Fig. 6), the section was cropped to the invasion front. Counts from each 2 × 2 group of source 8-µm bins were summed to form a 16-µm bin, and the source-bin coordinates were averaged. Each aggregated bin received the most frequent nonmissing periphery label among its source bins (tumor, 50-µm invasion band or outer tissue) and the first nonmissing macrophage-subtype label among those bins. Three programs were scored from z-scored inferred marker expression over bins where the SCATTER-estimated proportion of the relevant cell type exceeded 0.05: a tumor-invasive program (TGFBI, PERP, LAMC2, MMP7, CEACAM6, VIM) in tumor cells; a fibroblast desmoplastic program (COL1A1, COL1A2, FN1, MMP2, POSTN, SPARC, DCN); and a macrophage-SPP1 program (SPP1, C1QC, SELENOP). For each spatial bin, dT and dS denote the distances to the nearest tumor-region bin and outer-tissue bin, respectively. Signed distance is −dS within the tumor region, dT in the outer tissue and dT−dS in the invasion band. Distances retain the original X and Y coordinate units. The affine relationship between the original 8-µm capture-grid indices and these coordinates gives approximately 34.35 µm per coordinate unit. Gradients were calculated as Spearman correlations between program scores and signed distance across individual eligible spatial bins in the analyzed region. The tumor invasion, fibroblast desmoplastic and macrophage SPP1 analyses included 22,760, 19,236 and 11,987 spatial bins, respectively. The CD8 T-cell proportion–distance correlation in main Fig. 6k was also calculated across individual spatial bins. The curves in main Fig. 6d were restricted to the 2nd–88th percentile range of signed distances, with distance intervals containing fewer than 40 eligible spatial bins omitted. Within the invasion band, each mixed-gene signal was attributed across cell types using the inferred proportion of each type multiplied by its recovered expression. Contributions were normalized separately for each gene and therefore reflect both inferred proportion and expression. Putative cell–cell communication was scored for nine curated ligand–receptor pairs as the mean, over invasion-band bins, of the product of the sender type's recovered ligand expression and the receiver type's recovered receptor expression. This expression-based ranking does not demonstrate signaling activity[9]. The macrophage-SPP1 program was compared between bins with source-study SPP1-positive macrophage subtype labels (not used in SCATTER fitting) and bins in the source-study “other” macrophage category by a one-sided Mann–Whitney U test (SPP1-positive greater than other), with rank-biserial effect size.

# Supplementary references

1. Luo, Y., Zhao, X., Ye, K. & Meng, D. STINR: Deciphering Spatial Transcriptomics via Implicit Neural Representation. in *2025 IEEE/CVF Conference on Computer Vision and Pattern Recognition (CVPR)* 25930–25939 (IEEE, Nashville, TN, USA, 2025). doi:10.1109/CVPR52734.2025.02415.

2. Kleshchevnikov, V. *et al.* Cell2location maps fine-grained cell types in spatial transcriptomics. *Nat. Biotechnol.* **40**, 661–671 (2022).

3. Lopez, R. *et al.* DestVI identifies continuums of cell types in spatial transcriptomics data. *Nat. Biotechnol.* **40**, 1360–1369 (2022).

4. Wang, M.-G., Chen, L. & Zhang, X.-F. Dual decoding of cell types and gene expression in spatial transcriptomics with PANDA. *Nucleic Acids Res.* **52**, 12173–12190 (2024).

5. Tu, J.-J., Yan, H., Zhang, X.-F. & Lin, Z. Precise gene expression deconvolution in spatial transcriptomics with STged. *Nucleic Acids Res.* **53**, gkaf087 (2025).

6. Benjamini, Y. & Hochberg, Y. Controlling the False Discovery Rate: A Practical and Powerful Approach to Multiple Testing. *J. R. Stat. Soc. Ser. B Stat. Methodol.* **57**, 289–300 (1995).

7. Moran, P. A. P. NOTES ON CONTINUOUS STOCHASTIC PHENOMENA. *Biometrika* **37**, 17–23 (1950).

8. Türei, D., Korcsmáros, T. & Saez-Rodriguez, J. OmniPath: guidelines and gateway for literature-curated signaling pathway resources. *Nat. Methods* **13**, 966–967 (2016).

9. Dimitrov, D. *et al.* Comparison of methods and resources for cell-cell communication inference from single-cell RNA-Seq data. *Nat. Commun.* **13**, 3224 (2022).

10. Kim, N. *et al.* Single-cell RNA sequencing demonstrates the molecular and cellular reprogramming of metastatic lung adenocarcinoma. *Nat. Commun.* **11**, 2285 (2020).

11. Srivatsan, S. R. *et al.* Embryo-scale, single-cell spatial transcriptomics. *Science* **373**, 111–117 (2021).

12. Garcia-Alonso, L., Holland, C. H., Ibrahim, M. M., Turei, D. & Saez-Rodriguez, J. Benchmark and integration of resources for the estimation of human transcription factor activities. *Genome Res.* **29**, 1363–1375 (2019).

13. Janesick, A. *et al.* High resolution mapping of the tumor microenvironment using integrated single-cell, spatial and in situ analysis. *Nat. Commun.* **14**, 8353 (2023).

14. Singhal, V. *et al.* BANKSY unifies cell typing and tissue domain segmentation for scalable spatial omics data analysis. *Nat. Genet.* **56**, 431–441 (2024).